\documentclass[aps,amsmath,amssymb,preprintnumbers,groupedaddress,twocolumn]{revtex4}
\usepackage{graphicx,psfrag,hyperref}
\usepackage{amsmath,bbm,array,amsfonts,graphicx,wrapfig,lscape,float,mathtools,multirow,longtable,amssymb}
\usepackage[dvipsnames]{xcolor}

\newcommand{\be}{\begin{equation}}
\newcommand{\ee}{\end{equation}}
\newcommand{\beq}{\begin{equation}}
\newcommand{\beql}[1]{\begin{equation}\label{#1}}
\newcommand{\eeq}{\end{equation}}
\newcommand{\ba}{\begin{array}}
\newcommand{\ea}{\end{array}}
\newcommand{\bea}{\begin{eqnarray}}
\newcommand{\beal}[1]{\begin{eqnarray}\label{#1}}
\newcommand{\eea}{\end{eqnarray}}
\newcommand{\ben}{\begin{enumerate}}
\newcommand{\een}{\end{enumerate}}
\newcommand{\bean}{\begin{eqnarray*}}
\newcommand{\eean}{\end{eqnarray*}}
\newcommand{\eref}[1]{(\ref{#1})}
\newcommand{\sref}[1]{\S\ref{#1}}
\newcommand{\tref}[1]{Table~\ref{#1}}
\newcommand{\nn}{\nonumber}

\newcommand{\fref}[1]{Figure \ref{#1}}
\newcommand{\btab}[1]{\begin{tabular}{#1}}
\newcommand{\etab}{\end{tabular}}

\def \Black {\pdfsetcolor{0 0 0 1}}

\newcommand{\comment}[1]{}

\newcommand{\qed}{\nobreak \ifvmode \relax \else
      \ifdim\lastskip<1.5em \hskip-\lastskip
      \hskip1.5em plus0em minus0.5em \fi \nobreak
      \vrule height0.75em width0.5em depth0.25em\fi}

\newcommand{\Tr}{\text{Tr}}

\definecolor{darkspringgreen}{rgb}{0.09, 0.45, 0.27}
\definecolor{forestgreen}{rgb}{0.13, 0.55, 0.13}

\begin{document}

\preprint{UNIST-MTH-23-RS-04}
\title{Unsupervised Machine Learning Techniques for Exploring \\ Tropical Coamoeba,
Brane Tilings and Seiberg Duality}

\author{Rak-Kyeong Seong}

\email{seong@unist.ac.kr}

\affiliation{\it 
Department of Mathematical Sciences, and 
Department of Physics,\\ 
Ulsan National Institute of Science and Technology,\\
50 UNIST-gil, Ulsan 44919, South Korea
}

\begin{abstract}
We introduce unsupervised machine learning techniques in order to identify toric phases of $4d$ $\mathcal{N}=1$ supersymmetric gauge theories corresponding to the same toric Calabi-Yau 3-fold. 
These $4d$ $\mathcal{N}=1$ supersymmetric gauge theories are worldvolume theories of a D3-brane probing a toric Calabi-Yau 3-fold
and are realized in terms of a Type IIB brane configuration known as a brane tiling.
It corresponds to the skeleton graph of the coamoeba projection of the mirror curve associated to the toric Calabi-Yau 3-fold.
When we vary the complex structure moduli of the mirror Calabi-Yau 3-fold, the coamoeba and the corresponding brane tilings change their shape, giving rise to different toric phases related by Seiberg duality.
We illustrate that by employing techniques such as principal component analysis (PCA) and $t$-distributed stochastic neighbor embedding ($t$-SNE), we can project the space of coamoeba labelled by complex structure moduli down to a lower dimensional phase space with phase boundaries corresponding to Seiberg duality. 
In this work, we illustrate this technique by obtaining a 2-dimensional phase diagram for brane tilings corresponding to the cone over the zeroth Hirzebruch surface $F_0$.
\end{abstract} 
\maketitle
\noindent

\section{Introduction}

The worldvolume theories of a D3-brane probing a toric Calabi-Yau 3-fold \cite{fulton,1997hep.th...11013L} form a very rich class of $4d$ $\mathcal{N}=1$ supersymmetric gauge theories \cite{Greene:1996cy,Douglas:1997de,Douglas:1996sw,Lawrence:1998ja,Feng:2000mi,Feng:2001xr}.
These supersymmetric gauge theories are realized by a Type IIB brane configuration that takes the form of a bipartite periodic graph on a 2-torus.
Such bipartite graphs have been extensively studied in mathematics as \textit{dimers} \cite{2003math.....10326K,kasteleyn1967graph}, and are known as \textit{brane tilings} \cite{Franco:2005rj, Hanany:2005ve,Franco:2005sm} in string theory.
More recently, brane tilings have also been studied in relation to integrable systems \cite{Franco:2011sz,Goncharov:2011hp}, scattering amplitudes \cite{Franco:2012mm,Arkani-Hamed:2012zlh} and lattice gauge theories \cite{Razamat:2021jkx,Franco:2022ziy}.

Under T-duality, the D3-brane probing the toric Calabi-Yau 3-fold singularity becomes a D5-brane suspended between a NS5-brane wrapping a holomorphic curve $\Sigma$ \cite{Leung:1997tw,Hanany:1998it, Aganagic:1999fe}. 
The holomorphic curve is given by 
\beal{es00a01}
\Sigma ~:~ P(x,y) = 0 ~,~
\eea
where $P(x,y)$ is the \textit{Newton polynomial} in $x,y \in \mathbb{C}^*$ of the toric diagram $\Delta$ corresponding to the toric Calabi-Yau 3-fold. 
The Newton polynomial is defined as
\beal{es00a02}
P(x,y) = \sum_{(n_x,n_y) \in \Delta} c_{(n_x,n_y)} x^{n_x} y^{n_y} ~,~
\eea
where $(n_x,n_y)\in\mathbb{Z}^2$ are the coordinates of the vertices of the 2-dimensional convex lattice polygon $\Delta$, and $c_{(n_x,n_y)}\in \mathbb{C}^*$ are complex coefficients, which are \textit{complex structure moduli} of the mirror Calabi-Yau 3-fold \cite{Hori:2000ck,Hori:2000kt,mirrorbook}.

\begin{figure}[ht!!]
\begin{center}
\resizebox{\hsize}{!}{
\includegraphics[height=5cm]{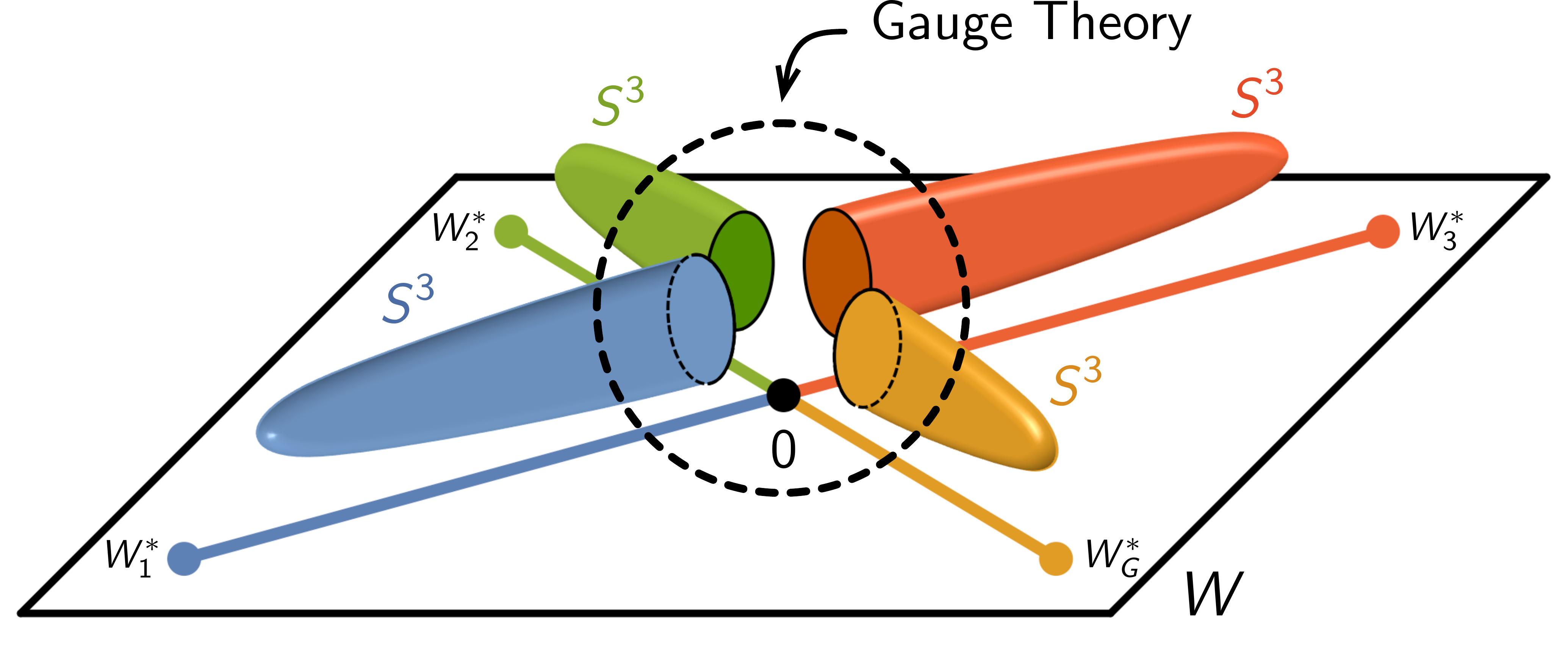} 
}
\caption{
The mirror geometry of a toric Calabi-Yau 3-fold is another 3-fold given by a double fibration over the complex $W$-plane.
Along a vanishing path connecting $W=0$ with a critical point $W^*$, we obtain from the double fibration an overall $S^3$.
The $S^3$'s meet at $W=0$ and how they intersect is given by the corresponding brane tiling that represents a $4d$ $\mathcal{N}=1$ supersymmetric gauge theory.
\label{f_fig01}}
 \end{center}
 \end{figure}

Brane tilings and the corresponding $4d$ $\mathcal{N}=1$ supersymmetric gauge theories have been extensively studied using \textit{mirror symmetry} of the corresponding toric Calabi-Yau 3-fold \cite{Hori:2000kt,Feng:2005gw}.
The mirror geometry of the toric Calabi-Yau 3-fold is another 3-fold given by the double fibration over the complex $W$-plane,
\beal{es00a02b}
W = P(x,y) ~,~~~
W= uv ~,~
\eea
where $u,v \in \mathbb{C}$.
The critical points of $P(x,y)$ are given by $(x^*,y^*)$ and satisfy
\beal{es1a06}
\frac{\partial}{\partial x} P(x,y) \Big|_{(x^*,y^*)} = 0 ~,~
\frac{\partial}{\partial y} P(x,y) \Big|_{(x^*,y^*)} = 0 ~.~
\eea
On the $W$-plane, the critical points correspond to $W^* = P(x^*,y^*)$.
When the toric diagram $\Delta$ contains at least one internal point, the number of critical points of $P$ matches the normalized area of the toric diagram, which is also the number of gauge groups $G$ in the corresponding $4d$ theory \cite{Franco:2016qxh}.

\begin{figure}[ht!!]
\begin{center}
\resizebox{\hsize}{!}{
\includegraphics[height=5cm]{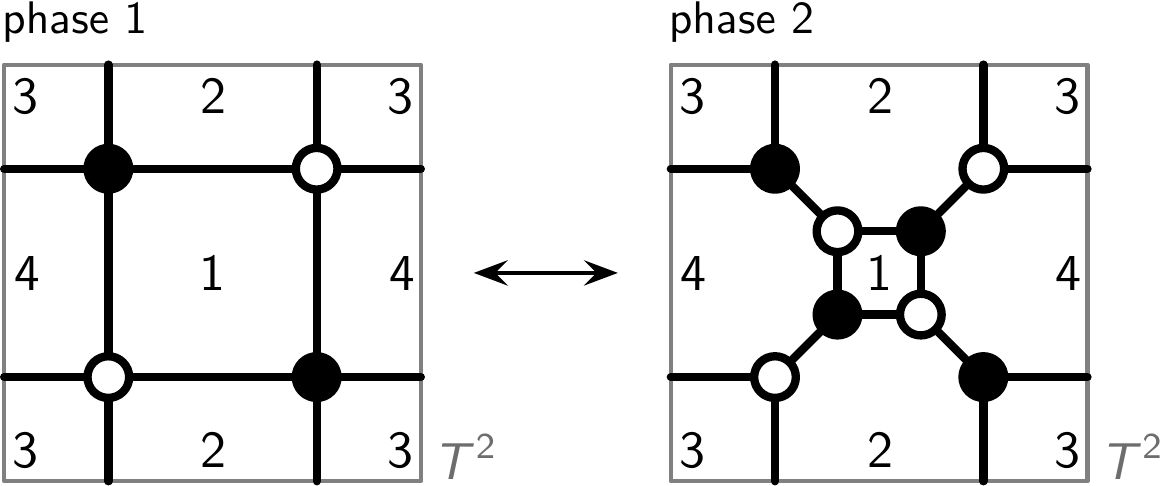} 
}
\caption{
A local deformation of a brane tiling known as urban renewal or spider move corresponds to Seiberg duality between the $4d$ $\mathcal{N}=1$ theories represented by the brane tilings connected by the local deformation.
\label{f_fig02}}
 \end{center}
 \end{figure}

The fiber associated to $P(x,y)$ in \eref{es00a02b} corresponds to a holomorphic curve $\Sigma_W: P(x,y)-W=0$, while the fiber associated to $uv$ in \eref{es00a02b} is a $\mathbb{C}^*$-fibration.
For generic values of the complex structure moduli $c_{(n_x,n_y)}$ in \eref{es00a02}, an $S^1\subset \Sigma_W$ shrinks to zero size at each critical point $W^*$, whereas additionally the $S^1$ from the $uv$-fibration vanishes at the origin $W=0$. 
Overall, from this $S^1 \times S^1$ fibration over a vanishing path connecting $W=0$ to a specific critical point $W^*$, we obtain an $S^3$ as illustrated in \fref{f_fig01}. 
All these $S^3$'s meet at $W=0$, where the $uv$-fibration vanishes. 
D6-branes wrapping these $S^3$'s give rise to the corresponding $4d$ $\mathcal{N}=1$ supersymmetric gauge theory, and the way how these spheres intersect each other at the vanishing locus $P(x,y)=W=0$ specifies the quiver and superpotential of the $4d$ theory. 
The intersection structure is precisely what is given by the holomorphic curve $\Sigma$ in \eref{es00a01}, which is represented by the corresponding brane tiling.

For different values of the complex structure moduli $c_{(n_x,n_y)}$, the positions of the critical points $W^*$ change and therefore the configuration of the $S^3$'s change.
This gives rise to a different intersection structure between the spheres and hence to a different $4d$ gauge theory phase. 
These phases are all represented by brane tilings and are also known as \textit{toric phases} \cite{Feng:2001bn,Feng:2000mi,Feng:2002zw}.
They are related by a local deformation of the brane tiling, which corresponds to Seiberg duality for $4d$ theories \cite{Seiberg:1994pq}.
This local deformation is also known as \textit{urban renewal} or \textit{spider move} \cite{ciucu1998complementation,Goncharov:2011hp,kenyon1999trees} in the mathematics literature and is illustrated in \fref{f_fig02}.

\begin{figure}[ht!!]
\begin{center}
\resizebox{\hsize}{!}{
\includegraphics[height=5cm]{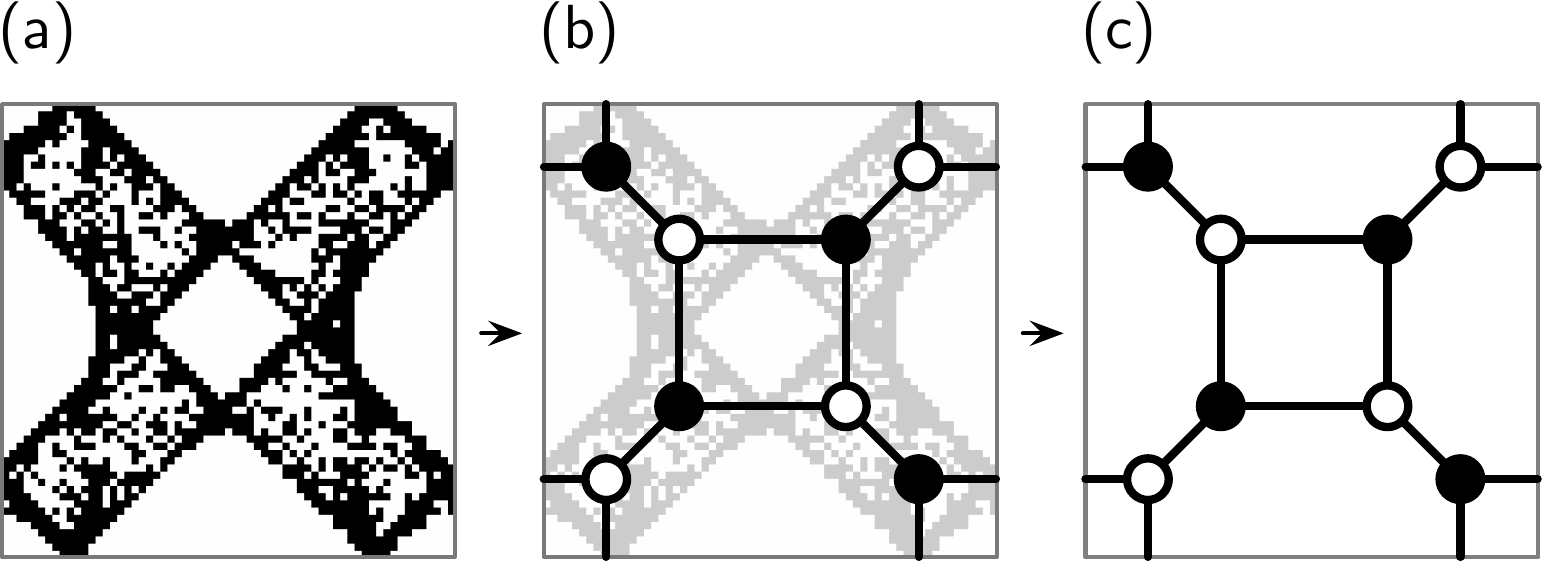} 
}
\caption{
(a) A coamoeba plot on $T^2$ (b) can be mapped to its corresponding unique skeleton graph, (c) which is a brane tiling on $T^2$.
\label{f_fig03}}
 \end{center}
 \end{figure}

In the following work, we focus on brane tilings given by the \textit{coamoeba} description \cite{2003math.ph..11005K,Feng:2005gw} of the holomorphic curve $\Sigma$. 
When we project the holomorphic curve $\Sigma$ onto a 2-torus using
\beal{es00a03}
(x,y) = (r_x e^{i\theta_x}, r_y e^{i\theta_y}) \mapsto (\theta_x, \theta_y) \in T^2 ~,~
\eea
we obtain what is known as the \textit{coamoeba projection} \cite{2003math.ph..11005K,Feng:2005gw} of $\Sigma$.
The coamoeba projection identifies the locations on $T^2$ where the D5-brane meets the NS5-brane wrapping $\Sigma$ in the T-dual description of the D3-brane probing the toric Calabi-Yau 3-fold. 
The skeleton graph of the coamoeba, as illustrated in \fref{f_fig03}, is precisely what we call as the brane tiling \cite{Franco:2005rj, Hanany:2005ve,Franco:2005sm}, which represents the $4d$ $\mathcal{N}=1$ supersymmetric gauge theory corresponding to the toric Calabi-Yau 3-fold. 

The choice of complex structure moduli $c_{(n_x,n_y)}\in \mathbb{C}^*$ in the Newton polynomial $P(x,y)$ determines the shape of the coamoeba on $T^2$. 
When the complex structure moduli pass critical values, the coamoeba transforms into a new shape, which corresponds to a new toric phase related by Seiberg duality to the original $4d$ theory. 

Although, it is well understood that the complex structure moduli $c_{(n_x,n_y)}\in \mathbb{C}^*$ affect the shape of the coamoeba, it is in general not well understood which values of the complex structure moduli correspond to which toric phase. 
We propose in this work a new method to explore and parameterize using complex structure moduli the phase space for brane tilings corresponding to a given toric Calabi-Yau 3-fold. 
Given that the number of complex structure moduli increases with increasing number of vertices in the toric diagram of the toric Calabi-Yau 3-fold, the problem of identifying regions in the space of all possible values of the complex structure moduli associated to specific toric phases is extremely challenging.

We therefore propose \textit{unsupervised machine learning} techniques in order to simplify 
the space of all possible coamoeba corresponding to a toric Calabi-Yau 3-fold. 
By using principal component analysis (PCA) \cite{pearson1901liii, hotelling1933analysis,jackson2005user,jolliffe2002principal,deisenroth2020mathematics} and additionally $t$-distributed stochastic neighbor embedding ($t$-SNE) \cite{hinton2002stochastic,van2008visualizing}, the space of coamoeba labelled by complex structure moduli can be dimensionally reduced to a lower dimensional phase space.
Such a phase space obtained using machine learning allows us to visualize regions in the phase space as toric phases and boundaries between regions as phase boundaries corresponding to Seiberg duality.  
Our proposed method also enables us to construct probability functions in terms of the complex structure moduli, whose probability values at certain choices of the complex structure moduli identify the corresponding toric phases.

Our work follows a long list of applications of machine learning techniques in string theory, beginning with the pioneering works in \cite{He:2017aed,Krefl:2017yox,Ruehle:2017mzq,Carifio:2017bov,Ruehle:2020jrk}, and aims for explainable and interpretable results in string theory and mathematics through machine learning.

\section{Brane Tilings and Coamoeba}\label{sback}

\begin{table}[ht!]
\begin{center}
\begin{tabular}{|c|cccc|cccc|cc|}
\hline
\; & $0$ & $1$ & $2$ & $3$ & $4$ & $5$ & $6$ & $7$ & $8$ & $9$ \\
\hline\hline
D5 & $\times$ & $\times$ & $\times$ & $\times$ & $\times$ & $\cdot$ & $\times$ & $\cdot$ & $\cdot$ & $\cdot$
\\
NS5 & $\times$ & $\times$ & $\times$ & $\times$ & \multicolumn{4}{c|}{--- $\Sigma$ ---} & $\cdot$ & $\cdot$
\\
\hline
\end{tabular}
\caption{
A $4d$ $\mathcal{N}=1$ gauge theory corresponding to a toric Calabi-Yau 3-fold can be represented by a Type IIB brane configuration known as a brane tiling.
\label{t_braneconfiguration}
}
\end{center}
\end{table}

$4d$ $\mathcal{N}=1$ supersymmetric gauge theories corresponding to toric Calabi-Yau 3-folds can be realized in terms of a Type IIB brane configuration known as a \textit{brane tiling} \cite{Franco:2005rj, Hanany:2005ve,Franco:2005sm}. 
A brane tiling consists of D5-branes suspended from a NS5-brane, where
the NS5-brane extends along the $(0123)$ directions and wraps a holomorphic curve $\Sigma$ embedded into the $(4567)$ directions. 
The coordinates $(45)$ and $(67)$ are pairwise combined into complex variables $x,y \in \mathbb{C}^*$, respectively.
They are the complex coordinates in the Newton polynomial $P(x,y)$ and the arguments $(\arg(x),\arg(y))=(\theta_x, \theta_y)$ are the coordinates of a $T^2$.
The brane configuration is summarized in \tref{t_braneconfiguration}.

\begin{figure}[ht!!]
\begin{center}
\resizebox{\hsize}{!}{
\includegraphics[height=5cm]{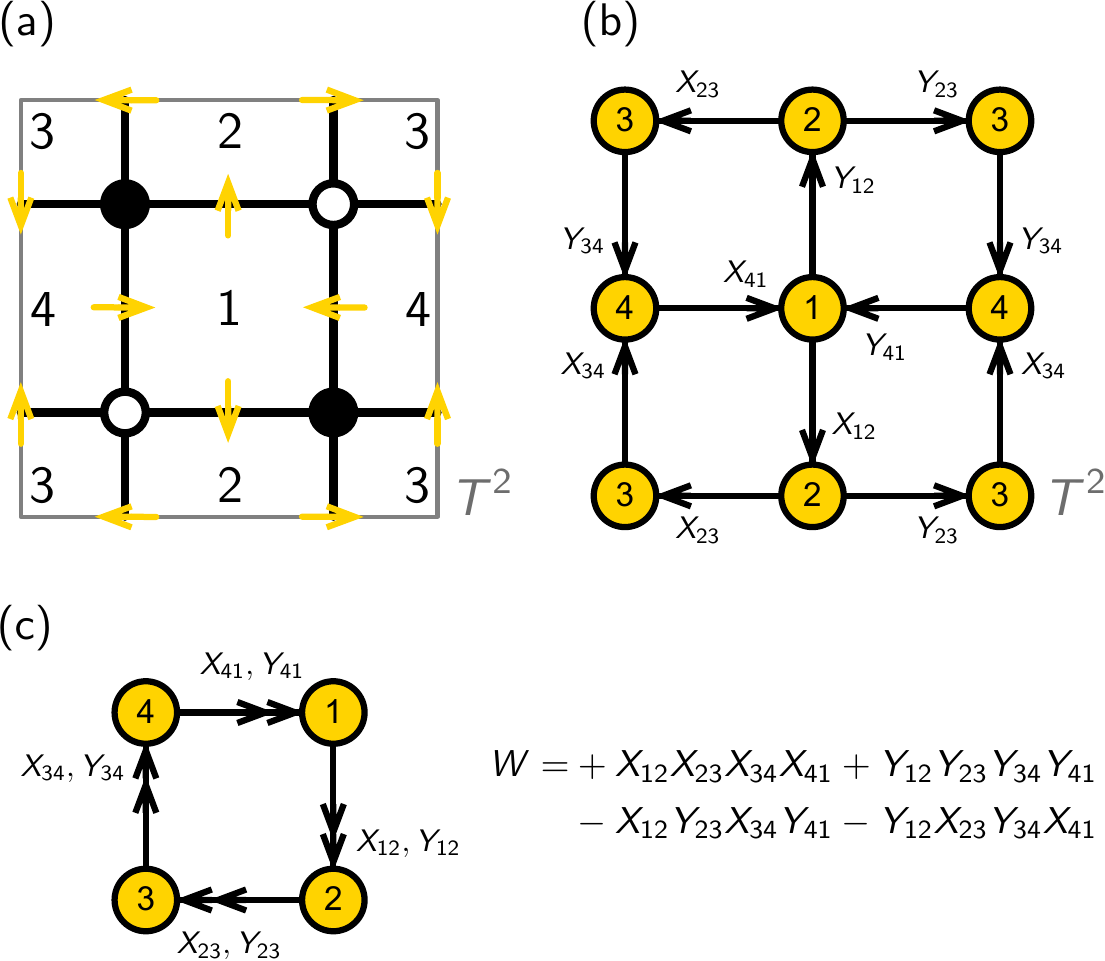} 
}
\caption{
(a) The brane tiling on $T^2$ with clockwise/anti-clockwise orientations on white/black nodes, (b) the periodic quiver on $T^2$, and (c) the quiver and superpotential corresponding to the first toric phase of $F_0$.
\label{f_fig04}}
 \end{center}
 \end{figure}

As discussed in \eref{es00a01}, the holomorphic curve $\Sigma$ is defined as the zero locus of the Newton polynomial $P(x,y)$ of the toric diagram $\Delta$. 
The brane tiling is the skeleton graph of the coamoeba projection of $\Sigma$ onto $(\arg(x),\arg(y)) \in T^2$ \cite{Feng:2005gw}. 
This is a projection of every point on the curve $\Sigma: P(x,y)=0$ to its angular component, resulting in a doubly periodic image of the curve $\Sigma$ on $T^2$. 
We refer to this image on $T^2$ as the \textit{coamoeba} \cite{2003math.ph..11005K,Feng:2005gw} of the toric Calabi-Yau 3-fold. 
The alternative projection of $\Sigma$ onto $(|x|, |y|) = (r_x,r_y)$ is called the \textit{amoeba} projection \cite{2004math......3015M,2001math......8225M}.

The brane tiling associated to the coamoeba is a bipartite periodic graph on $T^2$. 
We can use the following dictionary to identify the corresponding $4d$ $\mathcal{N}=1$ supersymmetric gauge theory from a brane tiling \cite{Franco:2005rj, Hanany:2005ve}:
\begin{itemize}
\item{\underline{White and black nodes.}} Positive and negative terms in the superpotential $W$ of the $4d$ $\mathcal{N}=1$ supersymmetric gauge theory correspond to white and black nodes in the brane tiling, respectively. 
The white and black nodes also have a clockwise and anti-clockwise orientation, respectively. 

\item{\underline{Edges.}} Bifundamental fields $X_{ij}$ in the $4d$ $\mathcal{N}=1$ supersymmetric gauge theory are represented by edges in the brane tiling, which connect always nodes that have different colors. 
Going along the orientation of a node, one can identify the fields $X_{ij}, X_{jk}, \dots X_{mi}$ associated to the specific superpotential term $\pm \Tr(X_{ij} X_{jk} \dots X_{mi}) \in W$ in the correct cyclic order.

\item{\underline{Faces.}} $U(N_i)$ gauge groups correspond to faces of a brane tiling. Every edge in the brane tiling corresponding to a bifundamental field $X_{ij}$ has neighboring faces associated to $U(N_i)$ and $U(N_j)$.
The orientation of the bifundamental field $X_{ij}$ is given by the orientation around the white and black nodes at the two opposite ends of the corresponding brane tiling edge. 
\end{itemize}
In the following work, all gauge groups are considered to be $U(1)$ such that the mesonic moduli space \cite{Witten:1993yc,Benvenuti:2006qr,Feng:2007ur,Butti:2007jv} of the $4d$ $\mathcal{N}=1$ supersymmetric gauge theories is precisely the probed toric Calabi-Yau 3-fold. 
Under Seiberg duality, dual brane tilings keep corresponding to abelian $4d$ theories with $U(1)$ gauge groups.
Accordingly, for abelian $4d$ theories corresponding to toric Calabi-Yau 3-folds, Seiberg duality is often referred to as \textit{toric duality} in the literature \cite{Feng:2001bn,Feng:2000mi,Feng:2001xr}.

As an example, \fref{f_fig04} shows the brane tiling for the first toric phase corresponding to the Calabi-Yau cone over the zeroth Hirzebruch surface $F_0$ \cite{brieskorn1966beispiele,hirzebruch1968singularities,Morrison:1998cs,Feng:2000mi}.
When we replace the faces as nodes, and edges as arrows along the orientation of the white and black nodes of the brane tiling, we obtain a \textit{periodic quiver} \cite{Franco:2005rj} on $T^2$.
The $T^2$ unit cell of the periodic quiver gives the \textit{quiver diagram} \cite{Douglas:1996sw} of the corresponding $4d$ $\mathcal{N}=1$ supersymmetric gauge theory.

\begin{figure}[ht!!]
\begin{center}
\resizebox{\hsize}{!}{
\includegraphics[height=5cm]{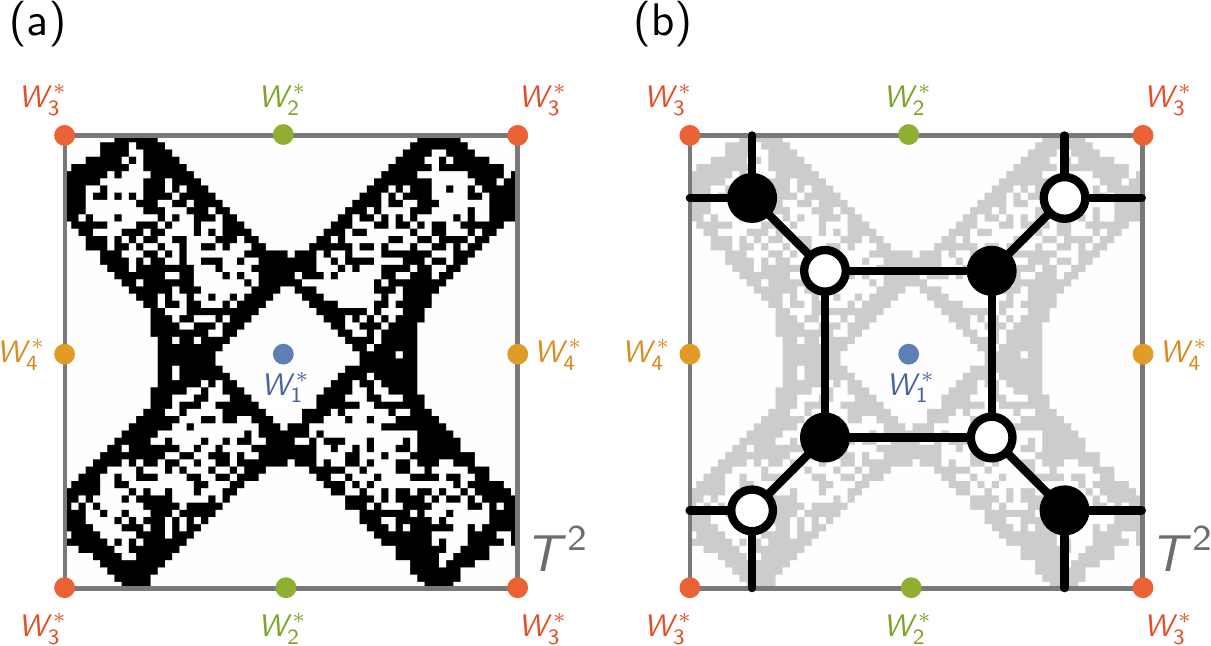} 
}
\caption{
Each critical point $W^*_i$ on the $W$-plane corresponds to a $U(N)$ gauge group in the $4d$ $\mathcal{N}=1$ theory. 
When they are projected under the coamoeba map to $T^2$, (a) they form the centers of white regions in the coamoeba plot and (b) are in the interior of the brane tiling faces. 
\label{f_fig05}}
 \end{center}
 \end{figure}

The faces of the brane tiling indicate the locations of the D5-brane suspended between the NS5-branes wrapping $\Sigma$ in the brane construction in \tref{t_braneconfiguration}.
In the mirror symmetry description, the faces of the brane tiling are centered around the critical points $W^*=(x^*,y^*)$ from \eref{es1a06}.
Each critical point $W^*_i$ corresponds to a gauge group, and when they are projected under the coamoeba map to $T^2$, they form the centers of the brane tiling faces.
The solutions to $P(x,y)=0$ mapped to $T^2$ indicate the location of the holomorphic curve $\Sigma$ on $T^2$ and is highlighted in black in the coamoeba plots as illustrated in \fref{f_fig05}.
These regions surround the critical points $W^*_i$ that are located inside the white regions of the coamoeba plot.
In the following section, we review how we generate these coamoeba plots and how we illustrate them on $T^2$.

\section{Coamoeba Generation and Representation}\label{scoamoeba}

For the following work, the coamoeba projection of $\Sigma$ onto $T^2$ plays a central role, 
because it directly corresponds to the brane tiling and $4d$ $\mathcal{N}=1$ supersymmetric gauge theory for a given toric Calabi-Yau 3-fold. 
From \eref{es00a02}, we note that $\Sigma: P(x,y)=0$ depends on the complex structure moduli $c_{(n_x,n_y)} \in \mathbb{C}^*$.
In order to identify a correspondence between complex structure moduli $c_{(n_x,n_y)}$ and specific toric phases, we have to generate for the same toric Calabi-Yau 3-fold many coamoeba for different values of $c_{(n_x,n_y)}$. 

As discussed above, the coamoeba is given by the angular projection of $\Sigma: P(x,y)=0$ onto $T^2$.
In order to obtain the coamoeba plot on $T^2$, one needs to solve for the set of solutions for $P(x,y)=0$, which then can be mapped to $T^2$ using the coamoeba map. 
We make use of the Monte-Carlo method \cite{metropolis1949monte} and scan for solutions in the range of $\theta_x = \arg(x) \in [0,2\pi)$ and $\theta_y = \arg(y) \in [0,2\pi)$. 
We call the process of searching for solutions while varying $\theta_x$ the $\theta_x$-scan, while the search for solutions while varying $\theta_y$ is called the $\theta_y$-scan.

For the $\theta_x$-scan, we first pick a random value $\theta_{x_0} \in [0,2\pi)$ and a random value for $m_\epsilon \in [- \epsilon,\epsilon]$, where the positive parameter $\epsilon \ll 1$ is taken to be small. 
Here, $\epsilon$ is a measure of the overall thickness of the coamoeba after projecting $\Sigma$ onto $T^2$.
By taking $x_0 = e^{i \theta_{x_0} + m_\epsilon}$, we can then numerically solve for $y$ using
\beal{es5a01}
P(x=x_0, y ) = \sum_{(n_x,n_y) \in \Delta} c_{(n_x,n_y)} x_0^{n_x} y^{n_y} = 0 ~.~
\eea
The solutions $y=y^*$ then combined with the value for $x_0$ can be mapped to $T^2$ by taking $(\theta_x^*, \theta_y^*)\equiv(\theta_{x_0}, \arg(y^*))$, where $\theta_{x_0}$ is given by $x_0 = e^{i \theta_{x_0} + m_\epsilon}$.
In total, we take $N_\theta$ random values for $\theta_{x_0} \in [0,2\pi)$ and $m_\epsilon \in [- \epsilon,\epsilon]$ in order to find corresponding solutions $y=y^*$ in \eref{es5a01}.
The resulting collection of solutions of the form $(\theta_{x_0}, \arg(y^*))$ gives the $\theta_x$-scan of the coamoeba. 
The $\theta_y$-scan of the coamoeba is similarly defined for a range of values for $\theta_y \in [0,2\pi)$.

\begin{figure}[ht!!]
\begin{center}
\resizebox{0.8\hsize}{!}{
\includegraphics[height=5cm]{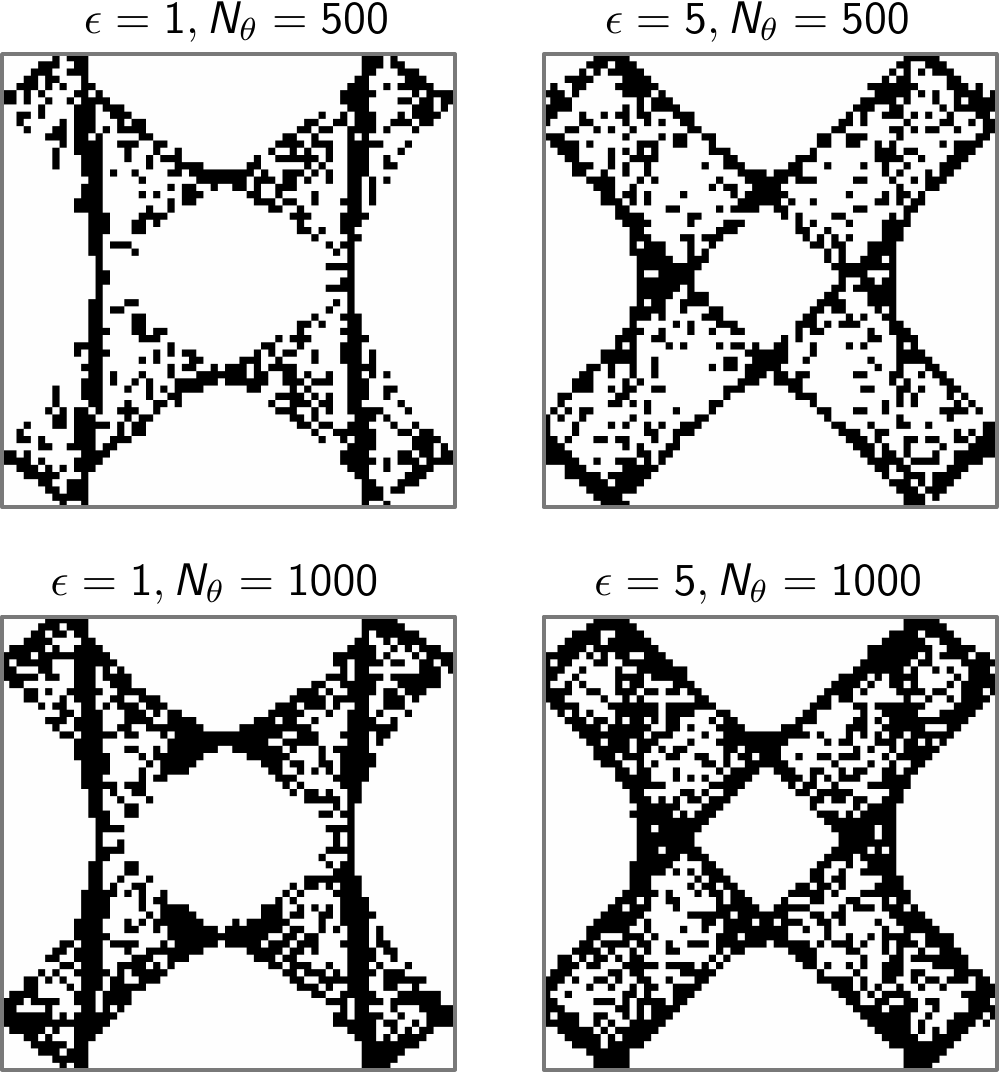} 
}
\caption{
Illustration of coamoeba plots corresponding to the cone over the zeroth Hirzebruch surface $F_0$ with the components of the complex structure moduli chosen to be $(c_{11}, c_{12}, c_{21}, c_{22})= (-3,-6,-3,+6)$.
The thickness parameter $\epsilon$ and the density parameter $N_\theta$ are varied between different plots.
\label{f_fig07}}
 \end{center}
 \end{figure}

\begin{figure*}[ht!!]
\begin{center}
\resizebox{0.98\hsize}{!}{
\includegraphics[height=5cm]{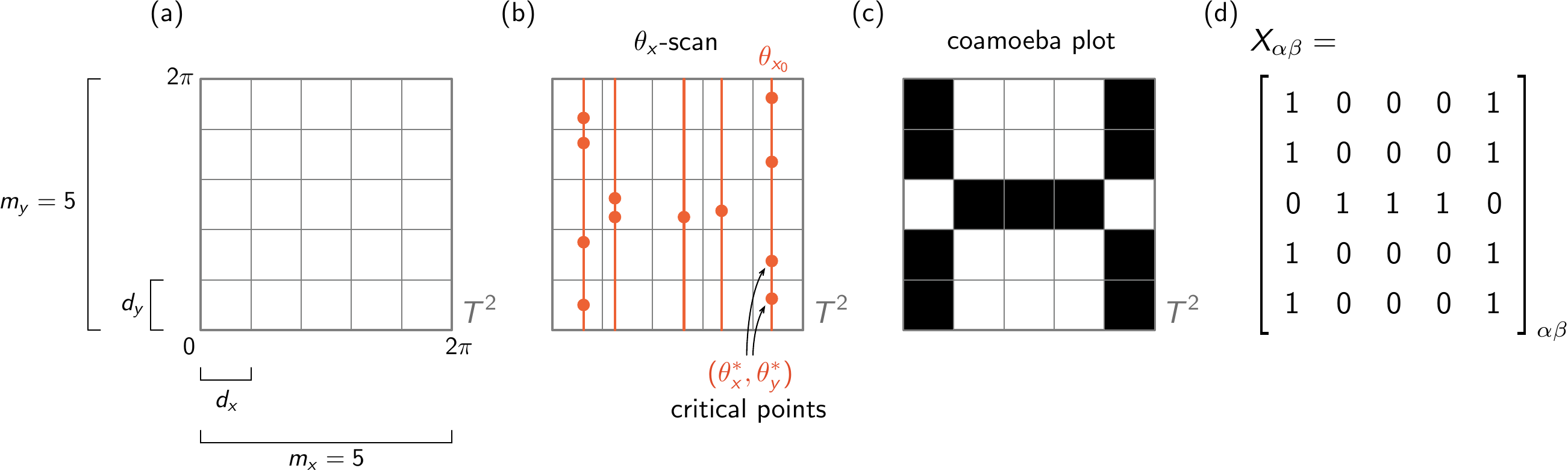} 
}
\caption{
A coamoeba plot consists of (a) a $m_x \times m_y$ grid with grid separation given by $d_x$ and $d_y$. (b) When a critical point $(\theta_x^*,\theta_y^*)$ is inside a grid point, (c) the grid point is colored black. (d) The grid can be represented in terms of a coamoeba matrix whose entries are $1$ for black grid points and otherwise $0$.
\label{f_fig06}}
 \end{center}
 \end{figure*}

For a given choice of complex structure parameters $c_{(n_x,n_y)} \in \mathbb{C}^*$ for a Newton polynomial $P(x,y)$ of a toric Calabi-Yau 3-fold, we define the components of the \textit{coamoeba matrix} $\textbf{X}$ as follows,
\beal{es5a10}
X_{\alpha\beta} = 
\left\{
\ba{llc}
1 & ~~ \exists ~(\theta_x^*, \theta_y^*)~ s.t.~ & (\beta-1) d_x \leq \theta_x^* < \beta d_x
\\
& ~~~~~~~~~~~~~~~ and & (\alpha-1) d_y \leq \theta_y^* < \alpha d_y
\\
0 & ~~ \text{otherwise}
\ea
\right.
~,~
\nn\\
\eea
where $(\theta_x^*, \theta_y^*)$ is a solution found under the $\theta_x$- and $\theta_y$-scans.
The indices $\alpha=1, \dots, m_y$ and $\beta=1,\dots,m_x$ label the grid points used for the coamoeba plot, and $0 < d_x \leq 2\pi$ and $0 < d_y \leq 2\pi$ measure the grid separation, as illustrated in \fref{f_fig06}(a).
The resolutions $m_x$ and $m_y$ for the coamoeba plot are defined by
\beal{es5a11}
d_x = \frac{2\pi}{m_x} 
~,~
d_y = \frac{2\pi}{m_y} 
~.~
\eea

Given the coamoeba matrix $\textbf{X}$, we can concatenate the rows of the matrix to form a $(m_x m_y)$-dimensional \textit{coamoeba vector} as follows,
\beal{es05a15}
\textbf{x} = (X_{11} , \dots , X_{1m_y}, X_{21}, \dots, X_{m_x 1}, \dots X_{m_x m_y})
~.~
\nn\\
\eea
These coamoeba vectors live in the positive cone of $\mathbb{Z}^{m_x m_y}$. 
We note that for each choice of complex structure moduli, we have a corresponding coamoeba vector in $\mathbb{Z}^{m_x m_y}$.
In the following work, we will use the coamoeba vectors as input to generate the phase diagram of coamoeba and brane tilings for a given toric Calabi-Yau 3-fold. 
\\

\noindent\textit{Example.} \fref{f_fig07} illustrates coamoeba plots corresponding to the cone over the zeroth Hirzebruch surface $F_0$, whose Newton polynomial takes the form,
\beal{es5a20}
P(x,y) = \left( x + \frac{1}{x} \right) + c_1 \left( y + \frac{1}{y} \right) + c_2
~,~
\eea
where $c_1,c_2 \in \mathbb{C}^*$ are the independent complex structure moduli.
We note that for any toric Calabi-Yau 3-fold, with a toric diagram $\Delta$ with $|\Delta|$ vertices on $\mathbb{Z}^2$, the complex coefficients in the Newton polynomial $P(x,y)$ can be rescaled such that they give precisely $|\Delta|-3$ independent ones \cite{Franco:2016qxh}. 
In the case for $F_0$, this precisely gives 2 independent complex structure moduli chosen to be $c_1=c_{11} + i c_{12}$ and $c_2=c_{21} + i c_{22}$ in \eref{es5a20}, where $c_{ij} \in \mathbb{R}$ such that $c_i \in \mathbb{C}^*$.
 \fref{f_fig07} shows the coamoeba plots for complex structure moduli set to $(c_{11}, c_{12}, c_{21}, c_{22}) = (-3.-6,-3,+6)$, with the coamoeba plot embedded in a grid defined by the resolution $m_x=m_y=63$, giving a total of $m_x m_y = 3969$
grid points in the coamoeba plots. 
Accordingly, the coamoeba vectors are in the positive cone of $\mathbb{Z}^{3969}$.
The four coamoeba plots in \fref{f_fig07} are obtained using $\epsilon \in \{1,5\}$ and $N_\theta=\{500,1000\}$ and are the union of the corresponding $\theta_x$- and $\theta_y$-scans.

\section{Dimensional Reduction and Principal Component Analysis}\label{spca}

Given a choice of complex structure moduli for a toric Calabi-Yau 3-fold, we can obtain through the coamoeba projection the corresponding coamoeba vector defined in \eref{es05a15}.
The choice of complex structure moduli and the corresponding coamoeba vector are both associated to a brane tiling representing a $4d$ $\mathcal{N}=1$ supersymmetric gauge theory.
Given that multiple choices of complex structure moduli can be associated to the same toric phase, identifying a map between toric phases and choices of complex structure moduli is a challenging problem.

In this work, we propose to dimensionally reduce the $(m_x m_y)$-dimensional coamoeba vectors in the positive cone of $\mathbb{Z}^{m_x m_y}$ to a lower dimensional phase space, where connected regions in the phase space can be identified with toric phases of brane tilings representing $4d$ $\mathcal{N}=1$ supersymmetric gauge theories. 
We use principal component analysis (PCA) \cite{pearson1901liii, hotelling1933analysis,jackson2005user,jolliffe2002principal}, which is one of the most fundamental and powerful techniques used for machine learning. 
The main idea behind PCA is to compress data by dimensional reduction with the aim of loosing the least amount of information during the process. 
In the following section, we give a brief review of PCA applied to coamoeba vectors relating to specific choices of complex structure moduli.
\\

\noindent\textit{Principal Component Analysis (PCA).}
Let us give a brief review of principal component analysis (PCA) \cite{pearson1901liii, hotelling1933analysis,jackson2005user,jolliffe2002principal,deisenroth2020mathematics} for coamoeba vectors associated to choices of complex structure moduli. 
Given a set of $N$ coamoeba vectors $\{ {\bf x}_1, \dots, { \bf  x}_N \}$, where for PCA we generalize ${\bf x}_a \in \mathbb{R}^d$ with $d= m_x m_y$, 
we can always project $\mathbb{R}^d$ to a lower $m$-dimensional feature space $U= \mathbb{R}^m \subset \mathbb{R}^d$. 
The lower dimensional representation of $\mathbf{x}_a \in \mathbb{R}^d$, which we call $\mathbf{z}_a = \pi_U(\mathbf{x}_a) \in U$, is given by
\beal{es40a00}
\mathbf{z}_a = \pi_U(\mathbf{x}_a) = P_\pi \mathbf{x}_a~,~
\eea
where the projection matrix from $\mathbb{R}^d$ to $U$ is given by 
\beal{es40a01}
P_\pi = \mathbf{B} (\mathbf{B}^\top \mathbf{B})^{-1} \mathbf{B}^\top 
~.~
\eea
Here, 
\beal{es40a02}
{\bf B}=\left[
{\bf b}_1, \dots, {\bf b}_m
\right] \in \mathbb{R}^{m\times d}~,~
\eea
is the matrix of basis vectors ${\bf b}_i$ for $U$.
The coordinate vector based on the basis given by ${\bf B}$ is $(z_{a1}, \dots, z_{am}) = ({\bf B}^\top {\bf B})^{-1} {\bf B}^\top  {\bf x}_a$.
Note that if the basis of $U$ is orthonormal, then ${\bf B}^\top {\bf B} = I$ gives the identity matrix, leading to
\beal{es40a05}
\mathbf{z}_a = \pi_U(\mathbf{x}_a) = \mathbf{B} \mathbf{B}^\top \mathbf{x}_a~,~
\eea
where 
\beal{es40a06}
(z_{a1}, \dots, z_{am}) = \mathbf{B}^\top \mathbf{x}_a~,~
\eea 
are the coordinates of $\mathbf{z}_a$ with respect to the orthonormal basis given by $\mathbf{B}$.

So far, we see that there are no surprises and using $P_\pi$ defined in terms of $\mathbf{B}$, we can project any coamoeba vector $\mathbf{x}_a\in \mathbb{R}^d$ into the subspace $U$.
This is however now the point where beautifully linear algebra meets statistics. 
What we mean is, we do not want to project $\mathbf{x}_a\in \mathbb{R}^d$ to any subspace $U$ of $\mathbb{R}^d$, but to a special subspace $\hat{U}$ where statistically the maximum amount of information in the original set of coamoeba vectors $\mathbf{x}_a\in \mathbb{R}^d$ is preserved. 
This in other words means we need to find $\hat{U}$ with basis vectors $\mathbf{b}_1, \dots, \mathbf{b}_m$ such that when $\mathbf{x}_a\in \mathbb{R}^d$ are projected to $\hat{U}$, they have a maximized variance along $\mathbf{b}_1, \dots, \mathbf{b}_m$. 

\begin{figure*}[ht!!]
\begin{center}
\resizebox{1\hsize}{!}{
\includegraphics[height=5cm]{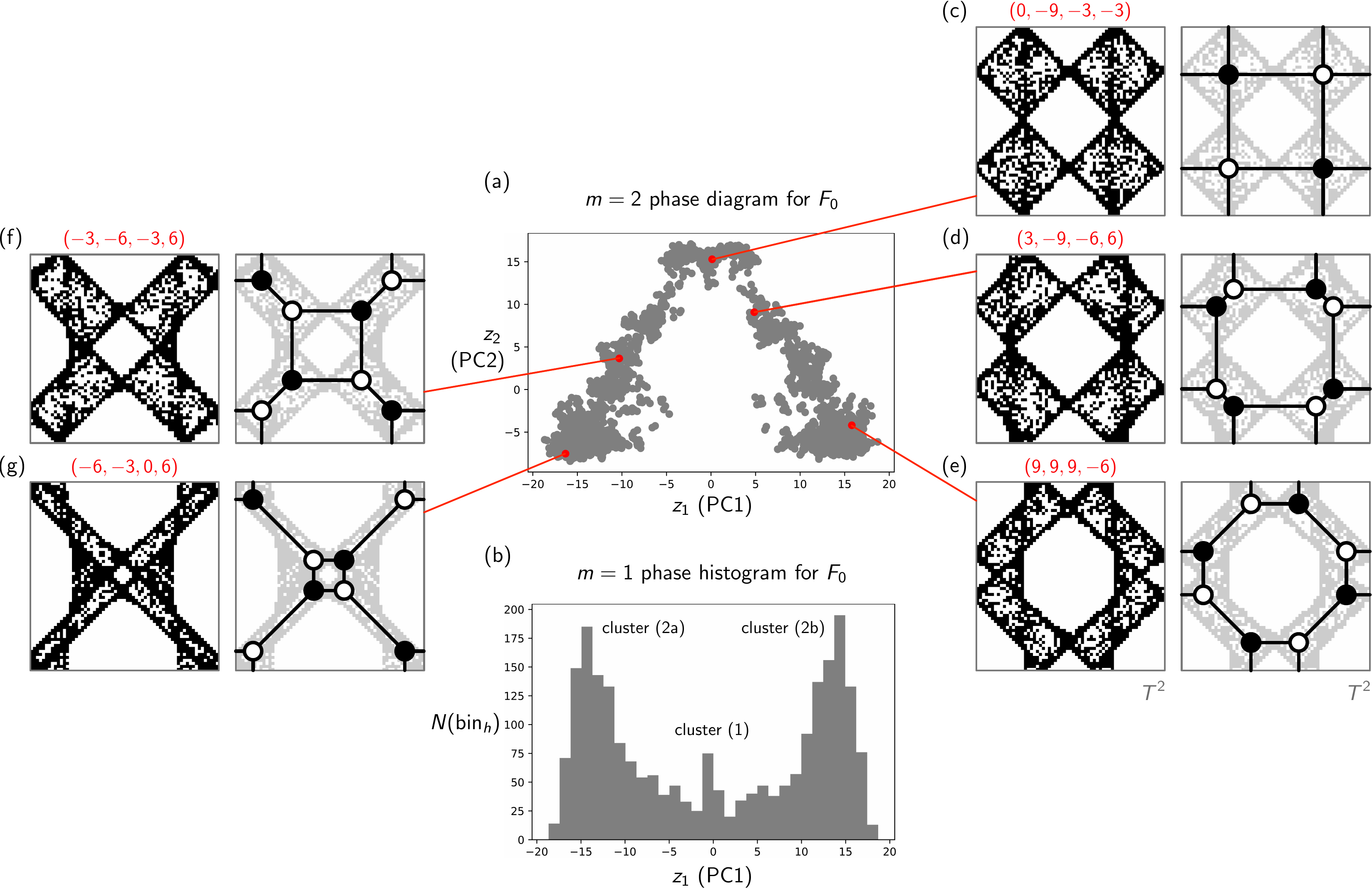} 
}
\caption{
(a) The $m=2$ phase diagram for $F_0$ consists of points that correspond to coamoeba plots and their corresponding brane tilings for specific choices of the complex structure moduli.
(b) We call the corresponding frequency histogram along the $z_1$-axis corresponding to the first principal component the $m=1$ phase histogram for $F_0$. It shows 3 peaks corresponding to 3 clusters in the $m=2$ phase diagram for $F_0$. We identify these clusters with 3 phases of coamoeba plots and their corresponding brane tilings, called phases (1), (2a) and (2b). 
In (c)-(g), we select 5 points in the $m=2$ phase diagram and illustrate the corresponding coamoeba plots, the associated brane tilings and the choices of complex structure moduli.
\label{f_fig08}}
 \end{center}
 \end{figure*}

Let us consider the first basis vector $\mathbf{b}_1$ of $\hat{U}$, that maximizes the variation of the original set of coamoeba vectors under the projection.
This involves the maximization of the variance of the first coordinate $z_{1a}$ of $\mathbf{z}_a$ over the whole set of vectors $\{ {\bf x}_1, \dots, { \bf  x}_N \}$.
Given that the basis of $\hat{U}$ is orthonormal, we can express the first coordinate as
\beal{es40a10}
z_{1a} = \mathbf{b}_1^\top \mathbf{x}_a
~.~ 
\eea
The \textit{variance} for the first coordinate $z_{1a}$ over $\{ \mathbf{x}_1, \dots, \mathbf{x}_N \}$ is then given by
\beal{es40a11}
V(z_1) = \frac{1}{N} \sum_{a=1}^{N} z_{1a}^2 = \frac{1}{N} \sum_{a=1}^{N} (\mathbf{b}_1^\top \mathbf{x}_{a})^2 ~,~
\eea
where we note that the expression for the variance is independent of the mean of the first coordinate over the whole set $\{ {\bf x}_1, \dots, { \bf  x}_N \}$.

Here, let us introduce the coamoeba \textit{covariance matrix} $\mathbf{S}$ for the original set of coamoeba vectors $\{ \mathbf{x}_1, \dots, \mathbf{x}_N \}$, which is defined as follows,
\beal{es40a12}
\mathbf{S}= \frac{1}{N} \sum_{a=1}^{N} \mathbf{x}_a \mathbf{x}_a^\top
~.~
\eea
The covariance matrix is symmetric and positive semidefinite and is a measure of how much the original set of coamoeba vectors is spread. 
The expression of the variance $V(z_1)$ of the first coordinate in \eref{es40a11} can be expressed in terms of the covariance matrix $\mathbf{S}$ as follows, 
\beal{es40a13}
V(z_1) = \mathbf{b}_1^\top \mathbf{S} \mathbf{b}_1 ~.~
\eea

The question now is how to maximize the variance $V(z_1)$. One option is to increase the length of the basis vector $\mathbf{b}_1$.
This is however not what we want given the fact that we started with the condition that $\hat{U}$ has an orthonormal basis. 
Accordingly, we introduce the constraint that the norm of the basis vector satisfies $|| \mathbf{b}_1||^2 = 1$. 
The result is a \textit{constrained optimization problem} taking the form,
\beal{es40a20}
&& \max_{\mathbf{b}_1} \left( V(z_1) \right) = \max_{\mathbf{b}_1} \left( \mathbf{b}_1^\top \mathbf{S} \mathbf{b}_1 \right) ~,~
\nn\\
&& ||\mathbf{b}_1 ||^2 = \mathbf{b}_1^\top \mathbf{b}_1 = 1 ~.~
\eea
In order to solve this constrained optimization problem, we write a Lagrangian function of the form
\beal{es40a22}
\mathcal{L}(\mathbf{b}_1, \lambda) =  \mathbf{b}_1^\top \mathbf{S} \mathbf{b}_1 + \lambda (1 - \mathbf{b}_1^\top \mathbf{b}_1 ) ~,~
\eea
where $\lambda$ is the Lagrange multiplier. 
By taking the partial derivatives of $\mathcal{L}(\mathbf{b}_1, \lambda)$ to zero, we obtain 
\beal{es40a23}
\mathbf{S} \mathbf{b}_1 = \lambda \mathbf{b}_1 ~,~
\eea
and $\mathbf{b}_1^\top \mathbf{b}_1 = 1$. 
Here, \eref{es40a23} beautifully refers to an eigenvalue equation, where the first basis vector $\mathbf{b}_1$ is the eigenvector with eigenvalue $\lambda$ for the coamoeba covariance matrix $\mathbf{S}$.

By inserting the eigenvalue equation in \eref{es40a23} with $\mathbf{b}_1^\top \mathbf{b}_1 = 1$ into the formula for the variance in \eref{es40a13}, 
we get
\beal{es40a25}
V(z_1) = \mathbf{b}_1^\top \mathbf{S} \mathbf{b}_1  = \lambda ~.~
\eea
This implies that in order to maximize the variance of the projected set of coamoeba vectors $\{ \mathbf{x}_1, \dots, \mathbf{x}_N \}$, we have to choose for $\mathbf{b}_1$ an eigenvector of the covariance matrix $\mathbf{S} $ that has the largest eigenvalue $\lambda=\lambda_1$.
This eigenvector $\mathbf{b}_1$ is also known as the first \textit{principal component}.
Using this principal component, we can then identify the \textit{optimal} projected coamoeba vector using \eref{es40a10} inside the original feature space $\mathbb{R}^d$, 
\beal{es40a26}
\hat{\mathbf{x}}_a = \mathbf{b}_1 z_{1a}  \in \mathbb{R}^d ~.~
\eea

We can now extend the problem to the $m$-th principal component.
Let us first assume that we already found the first $m-1$ principal components $\mathbf{b}_1, \dots, \mathbf{b}_{m-1}$, which correspond to the first $m-1$ eigenvectors of the coamoeba covariance matrix $\mathbf{S}$ with the eigenvalues sorted from large to small as follows $\lambda_1, \dots, \lambda_{m-1}$. 

For the $m$-th principal component, we then have to maximize the variance $V(z_m)$ for the $m$-th component of the set of coamoeba vectors $\{ {\bf x}_1, \dots, { \bf  x}_N \}$, 
under the constraint that we have orthonormal basis vectors satisfying $\mathbf{b}_m^\top \mathbf{b}_m =1$.
By solving the constrained optimization problem, 
we obtain the following eigenvalue equation 
\beal{es40a31}
\mathbf{S} \mathbf{b}_m = \lambda_m \mathbf{b}_m ~,~
\eea
where the variance of the coamoeba vectors projected onto the $m$-th principal component is given by the eigenvalue $V(z_m) = \lambda_m$.

Overall, in order to find the most optimal $m$-dimensional subspace $\hat{U}$ of the original space $\mathbb{R}^d$ of coamoeba vectors $\{ \mathbf{x}_1, \dots, \mathbf{x}_N \}$, we have to choose as basis vectors for $\hat{U}$ eigenvectors of the covariance matrix $\mathbf{S}$ that have the largest eigenvalues. 
The maximum variance from the first $m$ principal components is given by
\beal{es40a35}
V(z_1,\dots,z_m) = \sum_{i=1}^{m} \lambda_i ~.~
\eea

\noindent\textit{Unsupervised Optimization.}
The overall optimization problem of identifying the first $m$ principal components that maximize $V(z_1,\dots,z_m)$ can be reformulated as a \textit{loss function} $J(\mathbf{b}_a)$ that needs to be minimized. 
Such a loss function measures the overall difference between the actual coamoeba vectors $\mathbf{x}_a$ and the projected coamoeba vectors $\hat{\mathbf{x}}_a$ expressed in terms of $\mathbb{R}^d$ coordinates. 
Following the expression for $\mathbf{z}_a \in \hat{U}$ in \eref{es40a00}, we can express
\beal{es40a36}
\hat{\mathbf{x}}_a = (z_{1a}, \dots, z_{ma}, 0,\dots ,0) \in \mathbb{R}^d ~,~
\eea
where $(z_{1a}, \dots, z_{ma})$ are the coordinates for the projected coamoeba vector $\mathbf{z}_a$ in $\hat{U} \subset \mathbb{R}^d$.
Accordingly, we can write the loss function for the original set of coamoeba vectors $\{ \mathbf{x}_1, \dots, \mathbf{x}_N \}$ under PCA as
\beal{es40a40}
J(\mathbf{b}_a) 
&=& 
\frac{1}{N} \sum_{a=1}^N || \mathbf{x}_a - \hat{\mathbf{x}}_a ||^2 
\nn\\
&=&
\sum_{k=m+1}^d \mathbf{b}_k^\top \mathbf{S} \mathbf{b}_k
= \sum_{k=m+1}^d \lambda_{k} ~.~
\eea
We can see from here that in order to minimize the loss function above, we have to identify the smallest $d-m$ eigenvalues of the coamoeba covariance matrix $\mathbf{S}$ whose corresponding eigenvectors are orthogonal to the $m$ principal components that span $\hat{U}$.

In this work, we interpret $\hat{U}$ as a phase space for coamoeba corresponding to brane tilings and $4d$ $\mathcal{N}=1$ supersymmetric gauge theories associated to a toric Calabi-Yau 3-fold. 
Every projected vector $\mathbf{z}_a$ in $\hat{U}$ is a projected coamoeba vector with an associated choice of complex structure moduli.
If the dimension $m$ of $\hat{U}$ is kept small, we expect $\mathbf{z}_a$ to cluster into connected regions in $\hat{U}$ corresponding to coamoeba and brane tilings that form a toric phase associated to the toric Calabi-Yau 3-fold. 
In the following section, we obtain $\hat{U}$ for the Calabi-Yau cone over the zeroth Hirzebruch surface $F_0$ with $m=2$. 
We illustrate with this example that indeed $\hat{U}$ can be interpreted as a phase space for toric phases of $F_0$.
\\

\section{A Phase Diagram for Coamoeba and Brane Tilings}\label{sphase}

In this section, we overview the explicit construction of the phase diagram that we propose using PCA for coamoeba and brane tilings corresponding to the Calabi-Yau cone over the zeroth Hirzebruch surface $F_0$ \cite{brieskorn1966beispiele,hirzebruch1968singularities,Morrison:1998cs,Feng:2000mi}.
\\

\begin{figure}[ht!!]
\begin{center}
\resizebox{1\hsize}{!}{
\includegraphics[height=5cm]{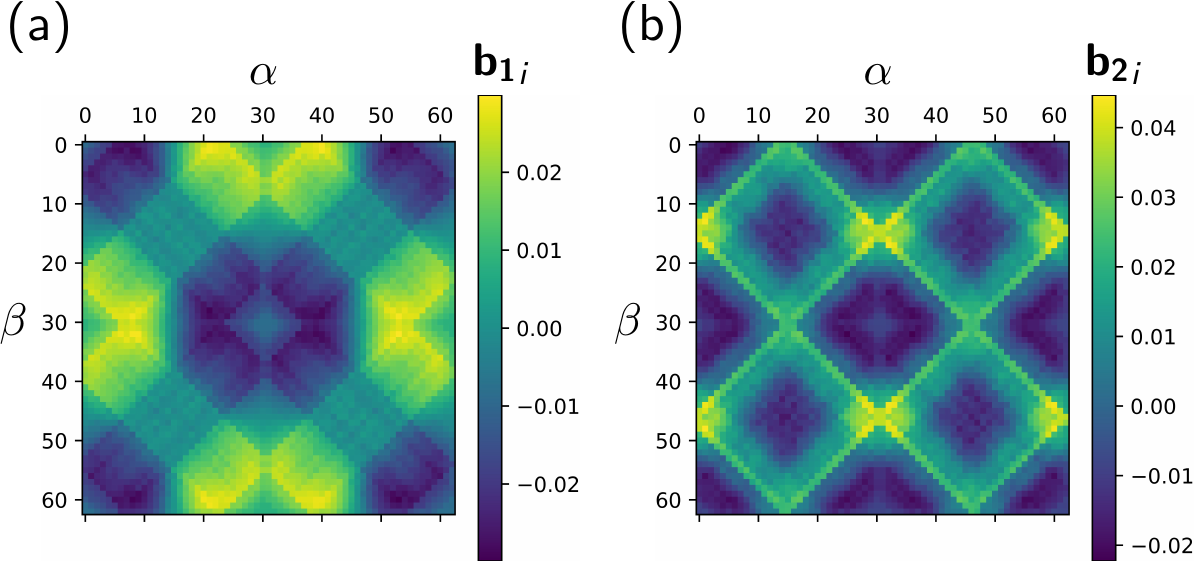} 
}
\caption{
The plots show the values of the $d=3969$ components of the eigenvectors (a) ${\bf b}_1$ and (b) ${\bf b}_2$ as heatmaps on the unit cell of $T^2$.
In (a), they form the shape of coamoeba corresponding to phase (2a) and (2b), whereas in (b), the heatmap forms the shape of the coamoeba for phase (1).
The indices $\alpha = 1, \dots, m_y$ and $\beta = 1, \dots, m_x$ label the grid points when the eigenvectors are illustrated in a $m_x \times m_y$ grid, analogous to coamoeba plots.
\label{f_fig09}}
 \end{center}
 \end{figure}

\noindent\textit{Example.}
Let us consider the brane tilings and coamoeba for the cone over the zeroth Hirzebruch surface $F_0$ \cite{brieskorn1966beispiele,hirzebruch1968singularities,Morrison:1998cs,Feng:2000mi}, whose Newton polynomial with complex structure moduli $c_1$ and $c_2$ is given in \eref{es5a20}.
Restricting ourselves to a discrete finite subset of all possible choices for the complex structure moduli, 
we choose the real moduli $c_{ij}$ to take values only from $\{-9,-6,-3,0,3,6,9\}$ such that $c_i \in \mathbb{C}^*$.
Under this restriction on the choices of complex structure moduli, we generate $N=2304$ coamoeba plots with parameters $\epsilon = 5$, $N_\theta = 2000$ and $m_x=m_y=63$. 
These give $N=2304$ coamoeba vectors ${\bf x}_a \in \mathbb{R}^d$ with $d=3969$.

We use PCA to project the coamoeba vectors ${\bf x}_a \in \mathbb{R}^d$ down to $\hat{U}$ with $m=2$ principal components ${\bf b}_1$ and ${\bf b}_2$.
These are the eigenvectors of the coamoeba covariance matrix $\textbf S$ defined in \eref{es40a12}, whose respective eigenvalues $\lambda_1$ and $\lambda_2$ are the largest and second largest out of all eigenvalues of $\textbf S$.
This means that along ${\bf b}_1, {\bf b}_2 \in \mathbb{R}^d$, the set of original coamoeba vectors ${\bf x}_a$ when projected down to ${\bf z}_a = (z_{1a},z_{2a}) \in \hat{U} \subset \mathbb{R}^d$, have the largest and second largest maximized variances $V(z_1)=\lambda_1$ and $V(z_2)=\lambda_2$ over all $a=1, \dots, N$.
The eigenvalues for our dataset of coamoeba vectors for $F_0$ take the following values,
\beal{es50a01}
\lambda_1 = 144.23
~,~
\lambda_2 = 57.79
~,~
\eea
where we express measurements up to 2 decimal points. 
The so-called \textit{proportional variance} given by the $i$-th eigenvalue of $\bf S$ can be calculated using
\beal{es50a10}
\lambda_i^* =\frac{\lambda_i}{\sum_{j=1}^{d} \lambda_j} \times 100\%
~,~
\eea
where the total variance $\sum_{j=1}^{d} \lambda_j$ is used in the denominator.
For our analysis, the proportional variance for the first $m=2$ principal components is given by
\beal{es50a11}
\lambda_1^* = 17.18 \%
~,~
\lambda_2^* = 6.88 \%
~.~
\eea
This means about $\lambda_1^* + \lambda_2^* = 24.06\%$ of the information in the original set of coamoeba vectors ${\bf x}_a$ is preserved in total if all the vectors are projected to ${\bf z}_a = (z_{1a},z_{2a}) \in \hat{U}$.

When we plot the coordinates $z_{1},z_{2}$ corresponding to the principal components ${\bf b}_1 (\text{PC1})$ and ${\bf b}_2 (\text{PC2})$, respectively,
for all coamoeba vectors, we obtain a 2-dimensional PCA score plot as shown in \fref{f_fig08}(a).
We interpret this plot as a 2-dimensional \textit{phase diagram} for brane tilings and coamoeba for $F_0$.
This is because the projected ${\bf z}_a$, each corresponding to a coamoeba defined for a particular choice of complex structure moduli whose skeleton graph is a brane tiling realizing a $4d$ $\mathcal{N}=1$ supersymmetric gauge theory, cluster in the 2-dimensional plot as shown in \fref{f_fig08}(a).

A closer look at the first principal component ${\bf b}_1 (\text{PC1})$ and the corresponding coordinate $z_1$ reveals that when plotted as a frequency histogram with bin size $\Delta z_1 = 1.25$ and with a total of $N_\text{bins}=30$ bins along the $z_1$-axis, there are 3 peaks in the frequency count $N(\text{bin}_h)$ of coamoeba vectors for a given $\text{bin}_h$ along the $z_1$-axis as shown in \fref{f_fig08}(b).
These peaks in the frequency count along the $z_1$-axis indicate clusters of projected coamoeba vectors ${\bf z}_a$ in the 2-dimensional phase diagram.
We call this frequency histogram along the $z_1$-axis corresponding to the first principal component the $m=1$ phase histogram for $F_0$. 

The locations of these clusters along the $z_1$-axis are given by
\beal{es50a12}
z_1^{\text{(1)}} = 0~,~ z_1^{\text{(2a)}} = -15 ~,~ z_1^{\text{(2b)}} = 15~,~
\eea
where we call the clusters respectively as (1), (2a) and (2b).
We note that the $m=2$ phase diagram appears to be symmetric around $z_1=0$, mapping clusters (2a) and (2b) into each other. 

When we take samples of projected vectors ${\bf z}_a$ from the phase diagram in \fref{f_fig08}(a), the corresponding coamoeba and their brane tilings, as shown in \fref{f_fig08}(c)-(g), contain the 2 known toric phases for the cone over the zeroth Hirzebruch surface $F_0$.
The first toric phase \cite{Morrison:1998cs,Feng:2001xr}, whose brane tiling, quiver diagram and superpotential of the corresponding $4d$ $\mathcal{N}=1$ supersymmetric gauge theory are shown in \fref{f_fig04}, can be associated to cluster (1) in the phase diagram in \fref{f_fig08}(a).
The second toric phase \cite{Feng:2000mi,Feng:2002zw}, which is Seiberg dual to the first toric phase, is identified with clusters (2a) and (2b), where as shown in \fref{f_fig08}(a) projected coamoeba vectors ${\bf z}_a$ in clusters (2a) and (2b) correspond to equivalent brane tilings on the 2-torus giving the same $4d$ $\mathcal{N}=1$ supersymmetric gauge theory.

We can further highlight the existence of these clusters by having a closer look at the eigenvectors ${\bf b}_1$ and ${\bf b}_2$ of the coamoeba covariance matrix $\bf S$.
We recall from \eref{es40a02} that these eigenvectors representing the two principal components form the matrix of basis vectors $\bf B$ of $\hat{U}$.
Following \eref{es40a06}, given that ${\bf B}^\top {\bf B} = I$ gives the identity matrix, ${\bf B}^\top$ projects the original coamoeba vectors ${\bf x_a}$ to the 2 coordinates $z_{1a}$ and $z_{2a}$.
An alternative interpretation of the eigenvectors ${\bf b}_1$ and ${\bf b}_2$ and the matrix ${\bf B}$ is that the absolute values of their components measure how correlated the $d=3969$ components of the original coamoeba vector ${\bf x_a}$ are with the $m=2$ components of the projected vectors with coordinates $(z_{1a},z_{2a})$ on the $m=2$ phase diagram.

We can in fact plot the values of the $d=3969$ components of the eigenvectors ${\bf b}_1$ and ${\bf b}_2$ on the unit cell of $T^2$, like the coamoeba plots obtained from the coamoeba matrix defined in \eref{es5a10}.
The resulting heatmaps for ${\bf b}_1$ and ${\bf b}_2$ are shown in \fref{f_fig09}.
Quite beautifully we can see that the large positive values (in yellow) for the components of ${\bf b}_1$ in \fref{f_fig09}(a) form the shape of a coamoeba corresponding to cluster (2a) in the $m=2$ phase diagram in \fref{f_fig08}(a).
Furthermore, the largely negative values (in dark blue) for the components of ${\bf b}_1$ form the shape of a coamoeba corresponding to cluster (2b) in the phase diagram in \fref{f_fig08}(a).
This means, the first principal component ${\bf b}_1$ mainly detects coamoeba corresponding to the second toric phase of $F_0$ and the sign of the components of ${\bf b}_1$ mainly distinguish between clusters (2a) and (2b) in the phase diagram in \fref{f_fig08}(a).
Moreover, looking at the heatmap for the second principal component ${\bf b}_2$ in \fref{f_fig09}(b) reveals that large positive values (in yellow) of the components of ${\bf b}_2$ form the shape of a coamoeba corresponding to cluster (1) in the phase diagram in \fref{f_fig08}(a).
This indicates that the second principal component ${\bf b}_2$ mainly identifies the first toric phase of $F_0$.
Accordingly, the 2 principal components ${\bf b}_1$ and ${\bf b}_2$ play a vital role in distinguishing coamoeba corresponding to the two toric phases of $F_0$.
\\

In our discussion, we have identified in the $m=2$ phase diagram for $F_0$ clusters of projected coamoeba vectors ${\bf z}_a$ by identifying peaks in the frequency histogram of vectors along the $z_1$-axis of the phase diagram.
In the following section, we employ further unsupervised machine learning techniques in order to explicitly identify the clusters and discrete boundaries between them. 
These boundaries can be interpreted as phase boundaries between toric phases of $F_0$ where Seiberg duality occurs.

\section{Clusters and Toric Phase Boundaries}\label{stsne}

\begin{figure*}[ht!!]
\begin{center}
\resizebox{0.72\hsize}{!}{
\includegraphics[height=5cm]{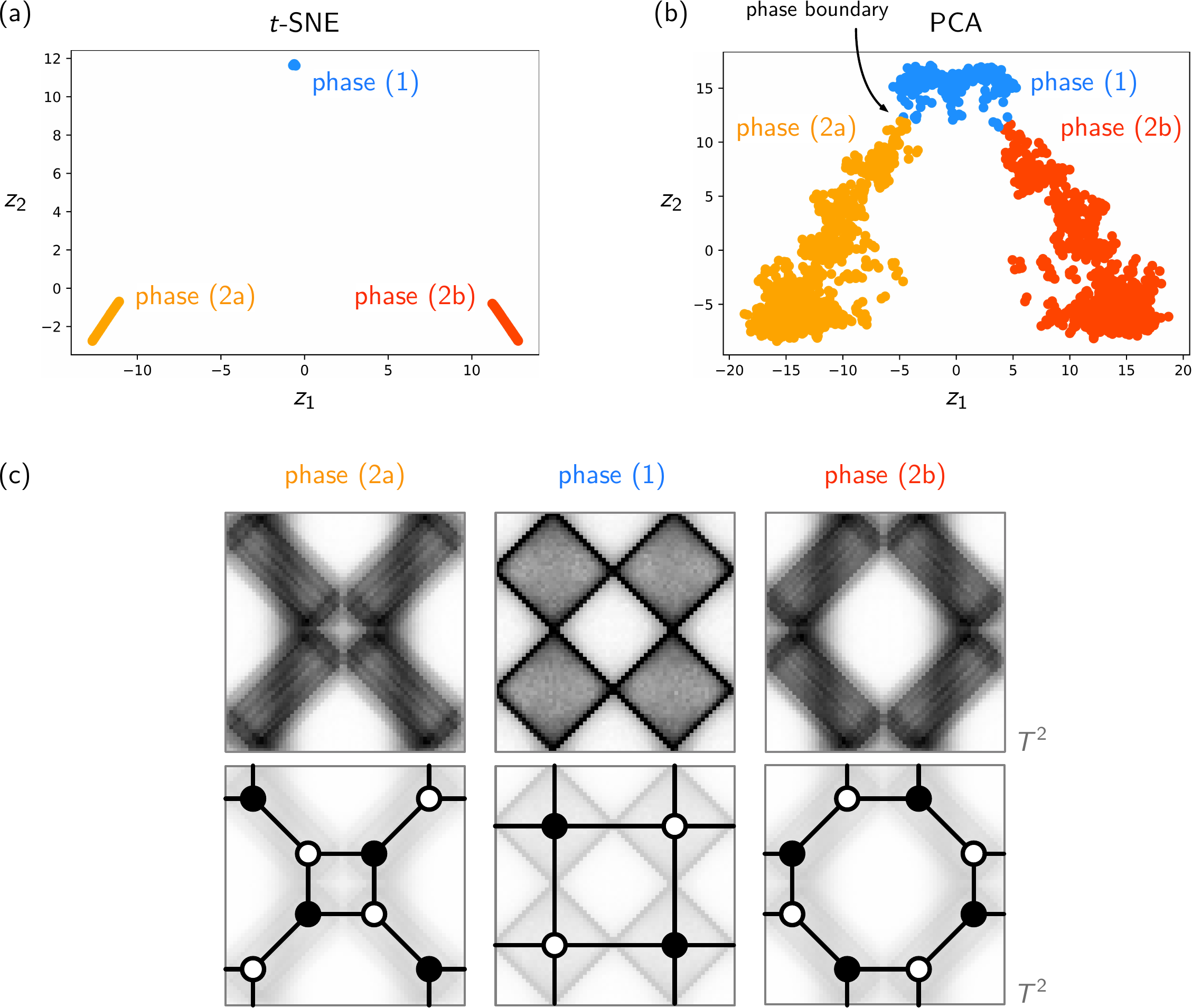} 
}
\caption{
(a) The $m=2$ $t$-SNE phase diagram for $F_0$ shows 3 clusters $S_p$ with sizes $|S_{\text{(2a)}}| = 1004$, $|S_{\text{(1)}}| = 297$, 
$|S_{\text{(2b)}}| = 1003$. The points corresponding to coamoeba for specific choices of complex structure moduli can be identified with points in the 
(b) $m=2$ PCA phase diagram for $F_0$. Given that through the $m=2$ $t$-SNE phase diagram we know which points correspond to which cluster and phase, we can identify the phase boundaries in the $m=2$ PCA phase diagram for $F_0$. In (c), we show the averaged coamoeba plots and the corresponding brane tilings for phases (1), (2a) and (2b).
\label{f_fig10}}
 \end{center}
 \end{figure*}

In order to identify the explicit borders between toric phases in the 2-dimensional phase diagram in \fref{f_fig08}(a), we make use of a manifold learning algorithm for dimensionally reducing multidimensional data known as \textit{$t$-distributed stochastic neighbor embedding} ($t$-SNE) \cite{hinton2002stochastic,van2008visualizing}.
Let us give a brief overview of $t$-SNE in terms of coamoeba vectors ${\bf x}_a \in \mathbb{R}^d$ with $a=1,\dots,N$.
\\

\noindent\textit{$t$-Distributed Stochastic Neighbor Embedding ($t$-SNE).}
Let us first define the \textit{pairwise affinity} $p_{b | a}$, which measures how similar two coamoeba vectors in the original $d$-dimensional space $\mathbb{R}^d$ are,
\beal{es60a01}
p_{b | a} = \frac{
\exp( - || {\bf x}_a - {\bf x}_b ||^2 / 2 \sigma_a^2)
}{
\sum_{h\neq a} \exp(- || {\bf x}_a - {\bf x}_h ||^2 / 2\sigma_a^2)
}
~.~
\eea
The pairwise affinity $p_{b | a}$ essentially measures the likelihood that one coamoeba vector ${\bf x}_a$ would pick ${\bf x}_b$ as its neighbor under the assumption that neighbors are picked following a Gaussian probability distribution centered at ${\bf x}_a$.
Here, the Gaussian is chosen to have a variance of $\sigma_a$. 
The pairwise affinity can be symmetrized as follows, 
\beal{es60a02}
p_{ab} = \frac{p_{b|a} + p_{a|b}}{2N} 
~,~
\eea
where $N$ is the number of coamoeba vectors ${\bf x}_a \in \mathbb{R}^d$.

We note that the individual variances $\sigma_a$ for each coamoeba vector ${\bf x}_a$ are determined using \textit{binary search}.
The search is dependent on the value of the \textit{perplexity} ${Perp}$ \cite{hinton2002stochastic,van2008visualizing}, which is defined as 
\beal{es60a05}
{Perp}(P_a) = 2^{H(P_a)}
~,~
\eea
where $P_a$ is the probability distribution formed by the conditional probabilities given by $p_{b | a}$ defined in \eref{es60a01}.
 $H(P_a)$ is the \textit{Shannon entropy} of $P_a$ given by,
\beal{es60a06}
H(P_a) = - \sum_{b=1}^N p_{b|a} \log_{2}{p_{b|a}} ~.~
\eea
Overall, the value of the perplexity ${Perp}$ has the effect of emphasizing local and global features of the dimensional reduction of the coamoeba vectors to ${\bf z}_a$ in $m$ dimensions. 

We can define a pairwise affinity between the dimensionally reduced coamoeba vectors ${\bf z}_a$ as follows,
\beal{es60a10}
q_{ab} = \frac{
(1+|| {\bf z}_a - {\bf z}_b ||^2)^{-1}
}{
 \sum_{h \neq g}^N (1+|| {\bf z}_h - {\bf z}_g ||)^{-1}
}
~,~
\eea
where here we use the $t$-distribution with $\nu=1$ degree of freedom. This gives us the Cauchy distribution which replaces the Gaussian distribution that is used in \eref{es60a01}.
Here, the pairwise affinity between the reduced coamoeba vectors ${\bf z}_a$ is a measure of similarity in the dimensionally reduced space.
The $t$-distribution is used for ${\bf z}_a$ rather than the Gaussian distribution, because the `heavy' tails of the distribution tend to make clusters of reduced coamoeba vectors ${\bf z}_a$ tighter than in other dimensional reduction techniques. 

Given the probability distribution $P_a$ defined by the pairwise affinities $p_{ab}$ of coamoeba vectors ${\bf x}_a \in \mathbb{R}^d$, 
and the probability distribution $Q_a$ defined by the pairwise affinities $q_{ab}$ of dimensionally reduced coamoeba vectors ${\bf z}_a \in \mathbb{R}^m$, 
we can measure the overall divergence of these two probability distributions using the \textit{Kullback-Leibler (KL) divergence} \cite{kullback1951information} defined as, 
\beal{es60a15}
C_{KL} = \sum_{a,b=1}^N p_{ab} \log \frac{p_{ab}}{q_{ab}}
~.~
\eea
The dimensional reduction of coamoeba vectors ${\bf x}_a \in \mathbb{R}^d$ to ${\bf z}_a \in \mathbb{R}^m$ using $t$-SNE has the aim of reducing the KL divergence $C$ by iteratively varying the positions of candidate coamoeba vectors ${\bf z}_a$ in $\mathbb{R}^m$.
This minimization process is done by gradient descent, where the gradient is defined as,
\beal{es60a20}
\frac{\delta C_{KL}}{\delta {\bf z}_a} = 4
\sum_{b=1}^{N} 
\frac{(p_{ab} - q_{ab}) ({\bf z}_a - {\bf z}_b) }{
1+ || {\bf z}_a - {\bf z}_b ||
} 
~.~
\eea

Having reviewed $t$-SNE for coamoeba vectors, let us now apply $t$-SNE to the $N=2304$ coamoeba vectors ${\bf x}_a \in \mathbb{R}^{3969}$ that we have obtained in section \sref{scoamoeba} for the Calabi-Yau cone over the zeroth Hirzebruch surface $F_0$.
\\

\noindent\textit{Example.}
Let us consider the $N=2304$ coamoeba vectors ${\bf x}_a \in \mathbb{R}^{3969}$ obtained for the cone over the zeroth Hirzebruch surface $F_0$ from section \sref{scoamoeba}.
Using $t$-SNE, the coamoeba vectors can be dimensionally reduced to ${\bf z}_a \in \mathbb{R}^m$ with $m=2$, giving the $t$-SNE phase diagram for $F_0$ shown in \fref{f_fig10}(a) with the perplexity parameter set to ${Perp} = 250$.
In comparison to the phase diagram obtained from PCA, the $t$-SNE phase diagram contains 3 disconnected clusters $S_p$, which can be identified with phases $p=\text{(1)}$, (2a) and (2b), where (2a) and (2b) correspond to the same second toric phase of $F_0$.

We can identify through the $t$-SNE phase diagram the coamoeba and the corresponding complex structure moduli corresponding to each of the 3 clusters $S_p$ in \fref{f_fig10}(a), giving the following counting of coamoeba,
\beal{es70a01}
|S_{\text{(2a)}}| = 1004 ~,~
|S_{\text{(1)}}| = 297 ~,~
|S_{\text{(2b)}}| = 1003 ~,~
\nn\\
\eea
where $N=|S_\text{(2a)}|+|S_\text{(1)}|+|S_\text{(2b)}|=2304$.

We define the \textit{average projected coamoeba vector} for a subset $S_p$ of vectors in a given phase $p$ as,
\beal{es70a05}
\langle {\bf z}_a \rangle_p = \frac{1}{|S_p|} \sum_{{\bf z}_a \in S_p} {\bf z}_a ~.~
\eea
Recalling that the concatenated rows of the coamoeba matrix give the coamoeba vector, we can illustrate the averaged coamoeba plots from the average coamoeba vectors $\langle {\bf z}_a \rangle_p$ for the 3 phases $S_p$ identified in the $t$-SNE phase diagram. 
These averaged coamoeba plots for each of the phases with the corresponding brane tilings are shown in \fref{f_fig10}(c).
We note that the brane tiling for the averaged coamoeba plot for phase $1$ corresponds exactly to the known first toric phase of $F_0$.
Moreover, the brane tilings for the averaged coamoeba plots for phases $2a$ and $2b$ corresponds exactly to the known second toric phase of $F_0$.

We can identify individual coamoeba vectors ${\bf z}_a$ in clusters $S_\text{(1)}$, $S_\text{(2a)}$ and $S_\text{(2b)}$ of the $m=2$ $t$-SNE phase diagram in the $m=2$ phase diagram obtained through PCA.
This allows us to identify the \textit{phase boundaries} between different phases in the $m=2$ PCA phase diagram as illustrated in \fref{f_fig10}(b).
We note that these phase boundaries in the $m=2$ phase diagram exactly correspond to where Seiberg duality occurs between toric phases. 

An interesting question one can ask is if one can predict now whether a certain choice of complex structure moduli in \eref{es5a20} leads to a specific toric phase of $F_0$.
We note that the clusters $S_\text{(1)}$, $S_\text{(2a)}$ and $S_\text{(2b)}$ obtained through the $t$-SNE phase diagram in \fref{f_fig10}(a) play an essential role in answering this question. 
In order to solve this problem, we propose the use of \textit{logistic regression} \cite{santner2012statistical,agresti2012categorical}, where the input parameters are chosen to be the components of the complex structure moduli $c_{11}$, $c_{12}$, $c_{21}$ and $c_{22}$ for $F_0$.
We recall that for the set of $N=2304$ coamoeba vectors, we chose values for $c_{ij} \in \{-9,-6,-3,0,3,6,9\}$ such that $c_i \in \mathbb{C}^*$ in section \sref{scoamoeba}.
For the purpose of using logistic regression, each component $c_{ij}$ is scaled as follows,
\beal{es70a10}
c_{ij}^\prime = \frac{c_{ij} - \mu(c_{ij})}{\sigma(c_{ij})}
~,~
\eea
where the mean $\mu(c_{ij})=0$ and standard deviation $\sigma(c_{ij}) = 6.06$ for all $c_{ij}$ over the $N=2304$ choices of complex structure moduli.

\begin{table}[ht!]
\begin{center}
\begin{tabular}{|p{10mm}|p{10mm}|p{10mm}|p{10mm}|p{10mm}|p{10mm}|p{10mm}|p{10mm}|}
\hline
$p$ & $|S_p|$ & $\beta^p_0$ & $\beta^p_1$ & $\beta^p_2$ & $\beta^p_3$ & $\beta^p_4$ 
\\
\hline\hline
(2a) & $1004$ & $-1.63$ & $-17.51$ & $+0.27$ & $-0.03$ & $+0.00$
\\
(1) & $297$ & $+3.20$ & $+0.00$ & $+0.00$ & $+0.22$ & $-0.03$
\\
(2b) & $1003$ & $-1.58$ & $+17.71$ & $-0.07$ & $+0.00$ & $+0.02$
\\
\hline
\end{tabular}
\caption{
Logistic regression result in terms of the real coefficients $\beta_i^p$ for the log odds of ${\bf z}_a$ belonging to cluster $S_p$ with complex structure moduli components $c_{ij}$.
The training was done on a random train ($80\%$) and test ($20\%$) set split of the $N=2304$ choices of complex structure moduli with an average test accuracy of $99.996 \%$. We can see from the coefficients $\beta_i^p$ that complex structure moduli component $c_{11}$ corresponding to $\beta_1^p$ plays the most significant role in determining the phase $p$ for ${\bf z}_a$.
\label{t_logreg}
}
\end{center}
\end{table}

In logistic regression, the aim is to obtain the probability that a given instance, in our case a choice of complex structure moduli given by $c_{ij}$, 
belongs to a particular class, in our case a particular subset of projected coamoeba vectors $S_p$.
The probability is given by,
\beal{es70a15}
P_p(c_{ij}) = \frac{1}{1- e^{- l_p(c_{ij})}}
~,~
\eea
where we choose $P_p(c_{ij}) = 1$ if the corresponding projected coamoeba vector ${\bf z}_a \in S_p$, and $P_p(c_{ij}) = 0$ if ${\bf z}_a \notin S_{p}$.
Here, logit - the \textit{log odds} of ${\bf z}_{a}$ belonging to $S_p$ with complex structure moduli components $c_{ij}$ -
is given by a linear combination of the complex structure moduli components, 
\beal{es70a16}
l_p(c_{ij}) = \beta^p_0 + \beta^p_1 c_{11} + \beta^p_2 c_{12} + \beta^p_3 c_{21} + \beta^p_4 c_{22}
~,~
 \nn \\
\eea
where $\beta^p_0 \in \mathbb{R}$ is the intercept and all other $\beta^p_i$ are real coefficients specific for a projected coamoeba vector ${\bf z}_a \in S_p$. 
In order to train the model, we randomly split the set of $N=2304$ choices of complex structure moduli that we use to generate the coamoeba vectors in section \sref{scoamoeba} into a train ($80\%$) and test ($20\%$) set.
With an average test accuracy of $99.996 \%$, we obtain the values for the intercept $\beta_0$ and coefficients $\beta_i$ for each of the projected coamoeba sets $S_p$ as summarized in \tref{t_logreg}.

We note that the absolute magnitude $|\beta_i^p|$ of the trained coefficients for a given set $S_p$ tells us about the strength of the relationship between the components of the complex structure moduli $c_{ij}$ in the logit function in \eref{es70a16} and the probability that the choice of complex structure moduli components $c_{ij}$ corresponds to a projected coamoeba vector ${\bf z}_a$ belonging to $S_p$ of phase $p$.
From the results in \tref{t_logreg}, we see that for phases (2a) and (2b) the complex structure moduli component $c_{11}$ with corresponding coefficient $\beta^p_1$ by far plays the greatest role in determining whether the projected coamoeba vector ${\bf z}_a$ corresponding to $c_{ij}$ is in $S_p$.
We recall that this is determined with an average test accuracy of $99.996 \%$.
A closer look reveals that
\beal{es80a01}
{\bf z}_a \in 
\left\{
\ba{cc}
S_\text{(1)}   & \text{if $c_{11} = 0$}\\
S_\text{(2a)} & \text{if $c_{11} < 0$}\\
S_\text{(2b)} & \text{if $c_{11} > 0$}\\
\ea
\right.
\eea
for any values of $c_{12}, c_{21}, c_{22} \in \mathbb{R}$ such that $c_i \in \mathbb{C}^*$.
\\

In summary, we identified using $t$-SNE the clusters of projected coamoeba vectors $S_\text{(2a)}$, $S_\text{(1)}$ and $S_\text{(2b)}$ corresponding to the toric phases of $F_0$.
We then provided evidence through logistic regression that just the sign of the component $c_{11}$ of the complex structure moduli for $F_0$ determines the phase of the corresponding coamoeba and brane tiling.
This is a remarkable result given that we started with a collection of coamoeba generated for an arbitrary range of complex structure moduli centered at 0 with no knowledge about which coamoeba and choice of complex structure moduli would correspond to which toric phase of $F_0$.

\section{Conclusions and Discussions}\label{sconc}

We made use of unsupervised machine learning techniques in order to find a phase diagram for the zeroth Hirzebruch surface $F_0$.
Such a phase diagram is useful because it identifies different
toric phases of $4d$ $\mathcal{N}=1$ supersymmetric gauge theories related by Seiberg duality in terms of
clusters in the phase diagram.
These clusters are made up of projected coamoeba vectors ${\bf z}_a$ which each refer to a coamoeba plot that corresponds to a brane tiling and a $4d$ $\mathcal{N}=1$ supersymmetric gauge theory.
The coamoeba plots are also associated to a specific choice of complex structure moduli in the mirror description of the toric Calabi-Yau 3-fold.
The phase diagram for $F_0$ was obtained by using PCA and $t$-SNE that projected the original coamoeba vectors in $d=3969$ dimensions to $2$ principal components corresponding to the axes of the $F_0$ phase diagram.
We even showed that $t$-SNE allows us to identify phase boundaries in the 2-dimensional phase diagram of $F_0$, where phase boundaries correspond to Seiberg duality between the $4d$ $\mathcal{N}=1$ supersymmetric gauge theories associated to the toric phases meeting at the phase boundary.
By the use of logistic regression, we gave evidence that amongst the complex structure moduli, only the value of a single real component of the moduli determines whether the corresponding coamoeba and brane tiling are in one of the 2 toric phases of $F_0$.

We expect that a similar construction of phase diagrams is possible for different toric Calabi-Yau 3-folds that exhibit a variety of toric phases in the context of the corresponding $4d$ $\mathcal{N}=1$ supersymmetric gauge theories \cite{Hanany:2012hi,Davey:2009bp}.
Moreover, we plan to report on phase diagrams constructed using the proposed unsupervised machine learning techniques for toric Calabi-Yau in higher dimensions, such as for toric Calabi-Yau 4-folds probed by D1-branes whose worldvolume theories are a class of $2d$ $(0,2)$ supersymmetric gauge theories realized in terms of brane brick models \cite{Franco:2016qxh,Franco:2015tya,Franco:2015tna}.
These $2d$ $(0,2)$ supersymmetric gauge theories related to toric Calabi-Yau 4-folds have been found to exhibit Gadde-Gukov-Putrov triality \cite{Gadde:2013lxa, Franco:2016nwv}
and are expected to have more elaborate phase diagrams that exhibit many interesting structures in the phase space of $2d$ $(0,2)$ theories.

The use of unsupervised machine learning techniques in this work in order to study the phase structure of supersymmetric gauge theories related to toric Calabi-Yau opens up a new avenue of research at the interface of supersymmetric gauge theories realized in string theory, Calabi-Yau mirror symmetry and tropical geometry, and explainable AI and unsupervised machine learning on which we hope to report on more in the near future.
\\

\acknowledgments

R.-K. S. is grateful to Per Berglund, Sebastian Franco, Sergei Gukov, Amihay Hanany, Yang-Hui He and Cumrun Vafa for discussions and feedback about this work. 
He would like to thank the Simons Center for Geometry and Physics at Stony Brook University for hospitality during various stages of this work.
He would also like to thank the organizers of the ``Dimers: Combinatorics, Representation Theory and Physics" workshop for their hospitality at the CUNY Graduate Center in New York. 
He is supported by a Basic Research Grant of the National Research Foundation of Korea (NRF-2022R1F1A1073128).
He is also supported by a Start-up Research Grant for new faculty at UNIST (1.210139.01), a UNIST AI Incubator Grant (1.230038.01) and UNIST UBSI Grants (1.230168.01, 1.230078.01), as well as an Industry Research Project (2.220916.01) funded by Samsung SDS in Korea.  
He is also partly supported by the BK21 Program (``Next Generation Education Program for Mathematical Sciences'', 4299990414089) funded by the Ministry of Education in Korea and the National Research Foundation of Korea (NRF).


\bibliographystyle{jhep}
\bibliography{mybib}

\providecommand{\href}[2]{#2}\begingroup\raggedright\begin{thebibliography}{10}

\bibitem{fulton}
W.~Fulton, \emph{{Introduction to toric varieties}}.
\newblock Annals of mathematics studies. Princeton Univ. Press, Princeton, NJ,
  1993.

\bibitem{1997hep.th...11013L}
N.~C. {Leung} and C.~{Vafa}, \emph{{Branes and Toric Geometry}}, {\emph{ArXiv
  High Energy Physics - Theory e-prints} (Nov., 1997) },
  [\href{http://arxiv.org/abs/hep-th/9711013}{{\tt hep-th/9711013}}].

\bibitem{Greene:1996cy}
B.~R. Greene, \emph{{String theory on Calabi-Yau manifolds}},  in
  \emph{{Theoretical Advanced Study Institute in Elementary Particle Physics
  (TASI 96): Fields, Strings, and Duality}}, pp.~543--726, 6, 1996.
\newblock \href{http://arxiv.org/abs/hep-th/9702155}{{\tt hep-th/9702155}}.

\bibitem{Douglas:1997de}
M.~R. Douglas, B.~R. Greene and D.~R. Morrison, \emph{{Orbifold resolution by
  D-branes}},
  \href{http://dx.doi.org/10.1016/S0550-3213(97)00517-8}{\emph{Nucl.Phys.} {\bf
  B506} (1997) 84--106}, [\href{http://arxiv.org/abs/hep-th/9704151}{{\tt
  hep-th/9704151}}].

\bibitem{Douglas:1996sw}
M.~R. Douglas and G.~W. Moore, \emph{{D-branes, Quivers, and ALE Instantons}},
  \href{http://arxiv.org/abs/hep-th/9603167}{{\tt hep-th/9603167}}.

\bibitem{Lawrence:1998ja}
A.~E. Lawrence, N.~Nekrasov and C.~Vafa, \emph{{On conformal field theories in
  four-dimensions}},
  \href{http://dx.doi.org/10.1016/S0550-3213(98)00495-7}{\emph{Nucl.Phys.} {\bf
  B533} (1998) 199--209}, [\href{http://arxiv.org/abs/hep-th/9803015}{{\tt
  hep-th/9803015}}].

\bibitem{Feng:2000mi}
B.~Feng, A.~Hanany and Y.-H. He, \emph{{D-brane gauge theories from toric
  singularities and toric duality}},
  \href{http://dx.doi.org/10.1016/S0550-3213(00)00699-4}{\emph{Nucl. Phys.}
  {\bf B595} (2001) 165--200}, [\href{http://arxiv.org/abs/hep-th/0003085}{{\tt
  hep-th/0003085}}].

\bibitem{Feng:2001xr}
B.~Feng, A.~Hanany and Y.-H. He, \emph{{Phase structure of D-brane gauge
  theories and toric duality}},
  \href{http://dx.doi.org/10.1088/1126-6708/2001/08/040}{\emph{JHEP} {\bf 08}
  (2001) 040}, [\href{http://arxiv.org/abs/hep-th/0104259}{{\tt
  hep-th/0104259}}].

\bibitem{2003math.....10326K}
R.~{Kenyon}, \emph{{An introduction to the dimer model}}, {\emph{ArXiv
  Mathematics e-prints} (Oct., 2003) },
  [\href{http://arxiv.org/abs/math/0310326}{{\tt math/0310326}}].

\bibitem{kasteleyn1967graph}
P.~Kasteleyn, \emph{Graph theory and crystal physics}, {\emph{Graph theory and
  theoretical physics} (1967) 43--110}.

\bibitem{Franco:2005rj}
S.~Franco, A.~Hanany, K.~D. Kennaway, D.~Vegh and B.~Wecht, \emph{{Brane Dimers
  and Quiver Gauge Theories}},
  \href{http://dx.doi.org/10.1088/1126-6708/2006/01/096}{\emph{JHEP} {\bf 01}
  (2006) 096}, [\href{http://arxiv.org/abs/hep-th/0504110}{{\tt
  hep-th/0504110}}].

\bibitem{Hanany:2005ve}
A.~Hanany and K.~D. Kennaway, \emph{{Dimer models and toric diagrams}},
  \href{http://arxiv.org/abs/hep-th/0503149}{{\tt hep-th/0503149}}.

\bibitem{Franco:2005sm}
S.~Franco et~al., \emph{{Gauge theories from toric geometry and brane
  tilings}}, \href{http://dx.doi.org/10.1088/1126-6708/2006/01/128}{\emph{JHEP}
  {\bf 01} (2006) 128}, [\href{http://arxiv.org/abs/hep-th/0505211}{{\tt
  hep-th/0505211}}].

\bibitem{Franco:2011sz}
S.~Franco, \emph{{Dimer Models, Integrable Systems and Quantum Teichmuller
  Space}}, \href{http://dx.doi.org/10.1007/JHEP09(2011)057}{\emph{JHEP} {\bf
  1109} (2011) 057}, [\href{http://arxiv.org/abs/1105.1777}{{\tt 1105.1777}}].

\bibitem{Goncharov:2011hp}
A.~B. Goncharov and R.~Kenyon, \emph{{Dimers and cluster integrable systems}},
  \href{http://arxiv.org/abs/1107.5588}{{\tt 1107.5588}}.

\bibitem{Franco:2012mm}
S.~Franco, \emph{{Bipartite Field Theories: from D-Brane Probes to Scattering
  Amplitudes}}, \href{http://dx.doi.org/10.1007/JHEP11(2012)141}{\emph{JHEP}
  {\bf 11} (2012) 141}, [\href{http://arxiv.org/abs/1207.0807}{{\tt
  1207.0807}}].

\bibitem{Arkani-Hamed:2012zlh}
N.~Arkani-Hamed, J.~L. Bourjaily, F.~Cachazo, A.~B. Goncharov, A.~Postnikov and
  J.~Trnka, \emph{{Grassmannian Geometry of Scattering Amplitudes}}.
\newblock Cambridge University Press, 4, 2016,
  \href{http://dx.doi.org/10.1017/CBO9781316091548}{10.1017/CBO9781316091548}.

\bibitem{Razamat:2021jkx}
S.~S. Razamat, \emph{{Quivers and Fractons}},
  \href{http://dx.doi.org/10.1103/PhysRevLett.127.141603}{\emph{Phys. Rev.
  Lett.} {\bf 127} (2021) 141603}, [\href{http://arxiv.org/abs/2107.06465}{{\tt
  2107.06465}}].

\bibitem{Franco:2022ziy}
S.~Franco and D.~Rodriguez-Gomez, \emph{{Quivers, Lattice Gauge Theories, and
  Fractons}},
  \href{http://dx.doi.org/10.1103/PhysRevLett.128.241603}{\emph{Phys. Rev.
  Lett.} {\bf 128} (2022) 241603}, [\href{http://arxiv.org/abs/2203.01335}{{\tt
  2203.01335}}].

\bibitem{Leung:1997tw}
N.~C. Leung and C.~Vafa, \emph{{Branes and toric geometry}},
  {\emph{Adv.Theor.Math.Phys.} {\bf 2} (1998) 91--118},
  [\href{http://arxiv.org/abs/hep-th/9711013}{{\tt hep-th/9711013}}].

\bibitem{Hanany:1998it}
A.~Hanany and A.~M. Uranga, \emph{{Brane boxes and branes on singularities}},
  \href{http://dx.doi.org/10.1088/1126-6708/1998/05/013}{\emph{JHEP} {\bf 9805}
  (1998) 013}, [\href{http://arxiv.org/abs/hep-th/9805139}{{\tt
  hep-th/9805139}}].

\bibitem{Aganagic:1999fe}
M.~Aganagic, A.~Karch, D.~Lust and A.~Miemiec, \emph{{Mirror symmetries for
  brane configurations and branes at singularities}},
  \href{http://dx.doi.org/10.1016/S0550-3213(99)00608-2}{\emph{Nucl. Phys. B}
  {\bf 569} (2000) 277--302}, [\href{http://arxiv.org/abs/hep-th/9903093}{{\tt
  hep-th/9903093}}].

\bibitem{Hori:2000ck}
K.~Hori, A.~Iqbal and C.~Vafa, \emph{{D-branes and mirror symmetry}},
  \href{http://arxiv.org/abs/hep-th/0005247}{{\tt hep-th/0005247}}.

\bibitem{Hori:2000kt}
K.~Hori and C.~Vafa, \emph{{Mirror symmetry}},
  \href{http://arxiv.org/abs/hep-th/0002222}{{\tt hep-th/0002222}}.

\bibitem{mirrorbook}
K.~Hori, S.~Katz, A.~Klemm, R.~Pandharipande, R.~Thomas, C.~Vafa et~al.,
  \emph{Mirror symmetry}, vol.~1 of \emph{Clay mathematics monographs}.
\newblock American Mathematical Society, Providence, RI, 2003.

\bibitem{Feng:2005gw}
B.~Feng, Y.-H. He, K.~D. Kennaway and C.~Vafa, \emph{{Dimer models from mirror
  symmetry and quivering amoebae}},
  \href{http://dx.doi.org/10.4310/ATMP.2008.v12.n3.a2}{\emph{Adv. Theor. Math.
  Phys.} {\bf 12} (2008) 489--545},
  [\href{http://arxiv.org/abs/hep-th/0511287}{{\tt hep-th/0511287}}].

\bibitem{Franco:2016qxh}
S.~Franco, S.~Lee, R.-K. Seong and C.~Vafa, \emph{{Brane Brick Models in the
  Mirror}}, \href{http://dx.doi.org/10.1007/JHEP02(2017)106}{\emph{JHEP} {\bf
  02} (2017) 106}, [\href{http://arxiv.org/abs/1609.01723}{{\tt 1609.01723}}].

\bibitem{Feng:2001bn}
B.~Feng, A.~Hanany, Y.-H. He and A.~M. Uranga, \emph{{Toric duality as Seiberg
  duality and brane diamonds}},
  \href{http://dx.doi.org/10.1088/1126-6708/2001/12/035}{\emph{JHEP} {\bf 12}
  (2001) 035}, [\href{http://arxiv.org/abs/hep-th/0109063}{{\tt
  hep-th/0109063}}].

\bibitem{Feng:2002zw}
B.~Feng, S.~Franco, A.~Hanany and Y.-H. He, \emph{{Symmetries of toric
  duality}}, \href{http://dx.doi.org/10.1088/1126-6708/2002/12/076}{\emph{JHEP}
  {\bf 12} (2002) 076}, [\href{http://arxiv.org/abs/hep-th/0205144}{{\tt
  hep-th/0205144}}].

\bibitem{Seiberg:1994pq}
N.~Seiberg, \emph{{Electric - magnetic duality in supersymmetric nonAbelian
  gauge theories}},
  \href{http://dx.doi.org/10.1016/0550-3213(94)00023-8}{\emph{Nucl. Phys.} {\bf
  B435} (1995) 129--146}, [\href{http://arxiv.org/abs/hep-th/9411149}{{\tt
  hep-th/9411149}}].

\bibitem{ciucu1998complementation}
M.~Ciucu, \emph{A complementation theorem for perfect matchings of graphs
  having a cellular completion}, {\emph{Journal of Combinatorial Theory, Series
  A} {\bf 81} (1998) 34--68}.

\bibitem{kenyon1999trees}
R.~W. Kenyon, J.~G. Propp and D.~B. Wilson, \emph{Trees and matchings},
  {\emph{arXiv preprint math/9903025} (1999) }.

\bibitem{2003math.ph..11005K}
R.~{Kenyon}, A.~{Okounkov} and S.~{Sheffield}, \emph{{Dimers and Amoebae}},
  {\emph{ArXiv Mathematical Physics e-prints} (Nov., 2003) },
  [\href{http://arxiv.org/abs/math-ph/0311005}{{\tt math-ph/0311005}}].

\bibitem{pearson1901liii}
K.~Pearson, \emph{Liii. on lines and planes of closest fit to systems of points
  in space}, {\emph{The London, Edinburgh, and Dublin philosophical magazine
  and journal of science} {\bf 2} (1901) 559--572}.

\bibitem{hotelling1933analysis}
H.~Hotelling, \emph{Analysis of a complex of statistical variables into
  principal components.}, {\emph{Journal of educational psychology} {\bf 24}
  (1933) 417}.

\bibitem{jackson2005user}
J.~E. Jackson, \emph{A user's guide to principal components}.
\newblock John Wiley \& Sons, 2005.

\bibitem{jolliffe2002principal}
I.~T. Jolliffe, \emph{Principal component analysis for special types of data}.
\newblock Springer, 2002.

\bibitem{deisenroth2020mathematics}
M.~P. Deisenroth, A.~A. Faisal and C.~S. Ong, \emph{Mathematics for machine
  learning}.
\newblock Cambridge University Press, 2020.

\bibitem{hinton2002stochastic}
G.~E. Hinton and S.~Roweis, \emph{Stochastic neighbor embedding},
  {\emph{Advances in neural information processing systems} {\bf 15} (2002) }.

\bibitem{van2008visualizing}
L.~Van~der Maaten and G.~Hinton, \emph{Visualizing data using t-sne.},
  {\emph{Journal of machine learning research} {\bf 9} (2008) }.

\bibitem{He:2017aed}
Y.-H. He, \emph{{Deep-Learning the Landscape}},
  \href{http://arxiv.org/abs/1706.02714}{{\tt 1706.02714}}.

\bibitem{Krefl:2017yox}
D.~Krefl and R.-K. Seong, \emph{{Machine Learning of Calabi-Yau Volumes}},
  \href{http://dx.doi.org/10.1103/PhysRevD.96.066014}{\emph{Phys. Rev. D} {\bf
  96} (2017) 066014}, [\href{http://arxiv.org/abs/1706.03346}{{\tt
  1706.03346}}].

\bibitem{Ruehle:2017mzq}
F.~Ruehle, \emph{{Evolving neural networks with genetic algorithms to study the
  String Landscape}},
  \href{http://dx.doi.org/10.1007/JHEP08(2017)038}{\emph{JHEP} {\bf 08} (2017)
  038}, [\href{http://arxiv.org/abs/1706.07024}{{\tt 1706.07024}}].

\bibitem{Carifio:2017bov}
J.~Carifio, J.~Halverson, D.~Krioukov and B.~D. Nelson, \emph{{Machine Learning
  in the String Landscape}},
  \href{http://dx.doi.org/10.1007/JHEP09(2017)157}{\emph{JHEP} {\bf 09} (2017)
  157}, [\href{http://arxiv.org/abs/1707.00655}{{\tt 1707.00655}}].

\bibitem{Ruehle:2020jrk}
F.~Ruehle, \emph{{Data science applications to string theory}},
  \href{http://dx.doi.org/10.1016/j.physrep.2019.09.005}{\emph{Phys. Rept.}
  {\bf 839} (2020) 1--117}.

\bibitem{2004math......3015M}
G.~{Mikhalkin}, \emph{{Amoebas of algebraic varieties and tropical geometry}},
  {\emph{ArXiv Mathematics e-prints} (Feb., 2004) },
  [\href{http://arxiv.org/abs/math/0403015}{{\tt math/0403015}}].

\bibitem{2001math......8225M}
G.~{Mikhalkin}, \emph{{Amoebas of algebraic varieties}}, {\emph{ArXiv
  Mathematics e-prints} (Aug., 2001) },
  [\href{http://arxiv.org/abs/math/0108225}{{\tt math/0108225}}].

\bibitem{Witten:1993yc}
E.~Witten, \emph{{Phases of N = 2 theories in two dimensions}},
  \href{http://dx.doi.org/10.1016/0550-3213(93)90033-L}{\emph{Nucl. Phys.} {\bf
  B403} (1993) 159--222}, [\href{http://arxiv.org/abs/hep-th/9301042}{{\tt
  hep-th/9301042}}].

\bibitem{Benvenuti:2006qr}
S.~Benvenuti, B.~Feng, A.~Hanany and Y.-H. He, \emph{{Counting BPS operators in
  gauge theories: Quivers, syzygies and plethystics}},
  \href{http://dx.doi.org/10.1088/1126-6708/2007/11/050}{\emph{JHEP} {\bf 11}
  (2007) 050}, [\href{http://arxiv.org/abs/hep-th/0608050}{{\tt
  hep-th/0608050}}].

\bibitem{Feng:2007ur}
B.~Feng, A.~Hanany and Y.-H. He, \emph{{Counting Gauge Invariants: the
  Plethystic Program}},
  \href{http://dx.doi.org/10.1088/1126-6708/2007/03/090}{\emph{JHEP} {\bf 03}
  (2007) 090}, [\href{http://arxiv.org/abs/hep-th/0701063}{{\tt
  hep-th/0701063}}].

\bibitem{Butti:2007jv}
A.~Butti, D.~Forcella, A.~Hanany, D.~Vegh and A.~Zaffaroni, \emph{{Counting
  Chiral Operators in Quiver Gauge Theories}},
  \href{http://dx.doi.org/10.1088/1126-6708/2007/11/092}{\emph{JHEP} {\bf 0711}
  (2007) 092}, [\href{http://arxiv.org/abs/0705.2771}{{\tt 0705.2771}}].

\bibitem{brieskorn1966beispiele}
E.~Brieskorn, \emph{Beispiele zur differentialtopologie von
  singularit{\"a}ten}, {\emph{Inventiones mathematicae} {\bf 2} (1966) 1--14}.

\bibitem{hirzebruch1968singularities}
F.~Hirzebruch, \emph{Singularities and exotic spheres}.
\newblock Societe Mathematic de France, 1968.

\bibitem{Morrison:1998cs}
D.~R. Morrison and M.~R. Plesser, \emph{{Nonspherical horizons. 1.}},
  {\emph{Adv.Theor.Math.Phys.} {\bf 3} (1999) 1--81},
  [\href{http://arxiv.org/abs/hep-th/9810201}{{\tt hep-th/9810201}}].

\bibitem{metropolis1949monte}
N.~Metropolis and S.~Ulam, \emph{The monte carlo method}, {\emph{Journal of the
  American statistical association} {\bf 44} (1949) 335--341}.

\bibitem{kullback1951information}
S.~Kullback and R.~A. Leibler, \emph{On information and sufficiency},
  {\emph{The annals of mathematical statistics} {\bf 22} (1951) 79--86}.

\bibitem{santner2012statistical}
T.~J. Santner and D.~E. Duffy, \emph{The statistical analysis of discrete
  data}.
\newblock Springer Science \& Business Media, 2012.

\bibitem{agresti2012categorical}
A.~Agresti, \emph{Categorical data analysis}, vol.~792.
\newblock John Wiley \& Sons, 2012.

\bibitem{Hanany:2012hi}
A.~Hanany and R.-K. Seong, \emph{{Brane Tilings and Reflexive Polygons}},
  \href{http://dx.doi.org/10.1002/prop.201200008}{\emph{Fortsch.Phys.} {\bf 60}
  (2012) 695--803}, [\href{http://arxiv.org/abs/1201.2614}{{\tt 1201.2614}}].

\bibitem{Davey:2009bp}
J.~Davey, A.~Hanany and J.~Pasukonis, \emph{{On the Classification of Brane
  Tilings}}, \href{http://dx.doi.org/10.1007/JHEP01(2010)078}{\emph{JHEP} {\bf
  01} (2010) 078}, [\href{http://arxiv.org/abs/0909.2868}{{\tt 0909.2868}}].

\bibitem{Franco:2015tya}
S.~Franco, S.~Lee and R.-K. Seong, \emph{{Brane Brick Models, Toric Calabi-Yau
  4-Folds and 2d (0,2) Quivers}},
  \href{http://dx.doi.org/10.1007/JHEP02(2016)047}{\emph{JHEP} {\bf 02} (2016)
  047}, [\href{http://arxiv.org/abs/1510.01744}{{\tt 1510.01744}}].

\bibitem{Franco:2015tna}
S.~Franco, D.~Ghim, S.~Lee, R.-K. Seong and D.~Yokoyama, \emph{{2d (0,2) Quiver
  Gauge Theories and D-Branes}},
  \href{http://dx.doi.org/10.1007/JHEP09(2015)072}{\emph{JHEP} {\bf 09} (2015)
  072}, [\href{http://arxiv.org/abs/1506.03818}{{\tt 1506.03818}}].

\bibitem{Gadde:2013lxa}
A.~Gadde, S.~Gukov and P.~Putrov, \emph{{(0, 2) trialities}},
  \href{http://dx.doi.org/10.1007/JHEP03(2014)076}{\emph{JHEP} {\bf 03} (2014)
  076}, [\href{http://arxiv.org/abs/1310.0818}{{\tt 1310.0818}}].

\bibitem{Franco:2016nwv}
S.~Franco, S.~Lee and R.-K. Seong, \emph{{Brane brick models and 2d (0, 2)
  triality}}, \href{http://dx.doi.org/10.1007/JHEP05(2016)020}{\emph{JHEP} {\bf
  05} (2016) 020}, [\href{http://arxiv.org/abs/1602.01834}{{\tt 1602.01834}}].

\end{thebibliography}\endgroup

\end{document}